\documentclass[11pt]{article}
\usepackage{cite}
\usepackage{amsmath,amsfonts,amssymb}%,calrsfs}
\usepackage{hyperref}
\usepackage[small,bf,hang]{caption}
\usepackage[usenames,dvipsnames]{xcolor}
\usepackage{mathabx}
\usepackage{latexsym,epsfig}
\usepackage{arydshln}

\def\hybrid{
        \topmargin -20pt
        \oddsidemargin 0pt
        \headheight 0pt \headsep 0pt
        \textwidth 6.25in % A4 paper
        \textheight 9.5in % A4 paper
        \marginparwidth .875in
        \parskip 5pt plus 1pt \jot = 1.5ex}

\hybrid

\renewcommand{\theequation}{\thesection.\arabic{equation}} \csname
@addtoreset\endcsname{equation}{section}

\def\moth{\mathsurround=0pt}
\newdimen\zo \zo=0pt

\def\tick{\leaders\hrule height 0.5ex depth 0pt \hskip 0.5pt}
\def\upboxfill{$\moth \setbox\zo\hbox{\tick}%
  \hskip 3pt\hbox to 0pt{$\tick$\hss}\hrulefill \hbox to 7.5pt{$\tick$\hss}$}
\def\underbox#1{\offinterlineskip{\mathord{\mathop{\vtop{\moth\ialign{##\crcr
      $\hfil\displaystyle{#1}\hfil$\crcr\noalign{}
      {\upboxfill}\crcr\noalign{}}}}\limits}}}
\def\dtick{\leaders\hrule height .34pt depth 0.5ex \hskip 0.5pt}
\def\downboxfill{$\moth \setbox\zo\hbox{\dtick}%
  \hskip 2pt\hbox to 0pt{$\dtick$\hss}\hrulefill \hbox to 2pt{$\dtick$\hss}$}
\def\overbox#1{\mathop{\vbox{\moth\ialign{##\crcr\noalign{}
\downboxfill\crcr\noalign{\vskip 1pt\nointerlineskip}
      $\hfil\displaystyle{#1}\hfil$\crcr}}}\limits}

\def\undersym#1{\underbox{{}#1}}
\def\oversym#1{\overbox{{}#1}}

\def\bec{\begin{center}}
\def\ec{\end{center}}

\def\L{\Lambda}
\def\m{\mu}

\def\cA{{\cal A}}

\def\cG{{\cal G}}

\def\cL{{\cal L}}
\def\cD{{\cal D}}
\def\cK{{\cal K}}
\def\cF{{\cal F}}

\def\cA{{\cal A}}

\def\cM{{\cal M}}
\def\cN{{\cal N}}
\def\cR{{\cal R}}
\def\cS{{\cal S}}
\def\cP{{\cal P}}
\def\cQ{{\cal Q}}

\def\cV{{\cal V}}

\def\cI{{\cal I}}
\def\cT{{\cal T}}

\def\cA{{\cal A}}

\def\be{\begin{equation}}
\def\ee{\end{equation}}
\def\bea{\begin{eqnarray}}
\def\eea{\end{eqnarray}}
\def\ba{\begin{array}}
\def\ea{\end{array}}

\usepackage{amssymb}
\usepackage{amsmath}
\usepackage{mathrsfs} %maths
\usepackage{youngtab}
\usepackage{mathtools}
\usepackage{stmaryrd}
\usepackage{bbm}
\usepackage{pifont}% 1,2,3 encercle
\newcommand{\R}{\mathcal{R}}%Ricci

\renewcommand{\L}{\mathbb{L}}%Dorfmann der
\renewcommand{\P}{\mathbb{P}}%projector
\selectfont

\begin{document}

\begin{titlepage}
\rightline{}
\rightline{\tt }
%\rightline{\tt  MIT-CTP-4604}
%\rightline\today
\rightline{July 2016}
\begin{center}
\vskip 1.0cm
{\Large \bf {E${}_{8(8)}$ Exceptional Field Theory:\\[1ex]
Geometry, Fermions and Supersymmetry}}\\
\vskip 1.0cm

     Arnaud Baguet, 
     Henning Samtleben
            \vskip .1cm
             \begin{small}
             			  {\it Univ Lyon, Ens de Lyon, Univ Claude Bernard, CNRS,\\
Laboratoire de Physique, F-69342 Lyon, France} \\
{{\tt arnaud.baguet, henning.samtleben@ens-lyon.fr}}
                        \end{small}

\vskip 1cm

\vskip 1cm
{\bf Abstract}
\end{center}

%\vskip 0.2cm

\begin{narrower}

\noindent
We present the supersymmetric extension of the recently constructed E$_{8(8)}$ exceptional field theory --
the manifestly U-duality covariant formulation of the untruncated ten- and eleven-dimensional
supergravities. This theory is formulated on a (3+248) dimensional spacetime (modulo section constraint) 
in which the extended coordinates transform in the adjoint representation of E$_{8(8)}$. 
All bosonic fields are E$_{8(8)}$ tensors and transform under internal generalized diffeomorphisms. 
The fermions are  tensors under the generalized Lorentz group ${\rm SO}(1,2)\times{\rm SO}(16)$, 
where SO(16) is the maximal compact subgroup of E$_{8(8)}$. 
Vanishing generalized torsion determines the corresponding spin connections to the extent they are 
required to formulate the field equations and supersymmetry transformation laws. 
We determine the supersymmetry transformations for all bosonic and fermionic fields such that they consistently 
close into generalized diffeomorphisms. In particular, the covariantly constrained gauge vectors of
E$_{8(8)}$ exceptional field theory
combine with the standard supergravity fields into a single supermultiplet.
We give the complete extended Lagrangian and show its invariance under supersymmetry.
Upon solution of the section constraint the theory reduces to full D=11 or type IIB supergravity.

\end{narrower}

\end{titlepage}
\newpage

\tableofcontents

%%%%%%%%%%%%%%%%%%%%%%%%%%%%%%%%%%%%%%%%%%%%%%
%%%%%%%%%%%%%%%%%%%%%%%%%%%%%%%%%%%%%%%%%%%%%%
\section{Introduction}
%%%%%%%%%%%%%%%%%%%%%%%%%%%%%%%%%%%%%%%%%%%%%%
%%%%%%%%%%%%%%%%%%%%%%%%%%%%%%%%%%%%%%%%%%%%%%

Exceptional field theories provide an ${\rm E}_{n(n)}$ covariant formulation of 
supergravity which unites eleven-dimensional supergravity and type IIB supergravity
in a common framework
\cite{Hohm:2013pua,Hohm:2013vpa,Hohm:2013uia,Hohm:2014fxa}.
This formulation gives a natural explanation for the exceptional symmetries known to
appear upon dimensional reduction of supergravity 
theories~\cite{Cremmer:1978ds,Cremmer:1979up,Cremmer:1980gs}.
It is based on the underlying symmetry algebra of ${\rm E}_{n(n)}$ generalized diffeomorphisms on an internal spacetime
whose coordinates are embedded into representations of the underlying exceptional 
groups (together with the associated section constraints)~\cite{Coimbra:2011ky,Berman:2012vc}. 
Action functionals invariant under
these generalized diffeomorphisms consistently reproduce subsectors of 
eleven-dimensional supergravity
\cite{Berman:2010is,Berman:2011jh,Coimbra:2011ky}.
Coupling of external tensor fields and further imposing invariance under diffeomorphisms on the external part of spacetime
determines a unique action functional -- the exceptional field theory -- which depending
on the solution of the section constraint reproduces 
the full eleven-dimensional supergravity and full type IIB supergravity, respectively.
For the lower rank groups ${\rm E}_{5(5)}={\rm D}_{5(5)}$, ${\rm E}_{4(4)}={\rm A}_{4(4)}$,
${\rm E}_{3(3)}={\rm A}_{2(2)}\otimes{\rm A}_{1(1)}$, and ${\rm E}_{2(2)}={\rm A}_{1(1)}\otimes\mathbb{R}^+$,
these theories have been constructed in 
\cite{Hohm:2015xna,Abzalov:2015ega,Musaev:2015ces,Berman:2015rcc}.
Remarkably, this construction uniquely reproduces the full bosonic sectors of the
higher-dimensional supergravities without any reference to the fermionic field content 
and supersymmetry. Nevertheless these actions can be supersymmetrized with fermions
transforming under the maximal compact subgroups ${\rm K}({\rm E}_{n(n)})$.
For ${\rm E}_{7(7)}$ and ${\rm E}_{6(6)}$ the supersymmetric completions have 
been worked out in \cite{Godazgar:2014nqa,Musaev:2014lna}.

The exceptional field theory based on the group ${\rm E}_{8(8)}$ appears to differ in some respects from
its lower-rank cousins. In the physical context the special role of ${\rm E}_{8(8)}$ 
is often assigned to the `dual graviton barrier'. Upon dimensional reduction to three dimensions, the
Kaluza-Klein vectors are dualized into scalar fields in order to exhibit the full duality group. As a result,
the scalar sector of the dimensionally reduced theory carries degrees of freedom descending from the
higher dimensional `dual graviton'~\cite{Curtright:1980yk,Hull:2000zn,West:2001as,Hull:2001iu}.
While this dualization goes through straightforwardly for the free vector fields of the dimensionally reduced theory,
their non-abelian gauge structure in presence of the higher Kaluza-Klein modes requires a modification
of the standard dualization procedure. On the formal side, this is reflected by the fact that the ${\rm E}_{8(8)}$
generalized diffeomorphisms do not close into an algebra~\cite{Coimbra:2011ky,Berman:2012vc} which has
obstructed a straightforward extension of the constructions for the lower-rank groups.
In E$_{8(8)}$ exceptional field theory this is taken care of 
by the appearance of an extra gauge symmetry in the commutator of two
generalized diffeomorphisms, such that the combined algebra closes and
allows for the construction of a gauge invariant action functional
which in turn reproduces the full higher-dimensional supergravities~\cite{Hohm:2014fxa}.
In particular, the realization of this extra gauge symmetry 
takes the form of constrained E$_{8(8)}$ rotations and
requires the introduction of an additional
(constrained) gauge connection $B_{\mu\,{\cal M}}$ which is invisible in the dimensionally reduced theory.
This extra gauge symmetry is in fact a generic feature of exceptional field theories but 
for the lower rank groups it only kicks in at the higher-rank $p$-forms. 
More specifically, the theories associated to the  E$_{n(n)}$ groups
exhibit such extra constrained gauge symmetry among the $(9-n)$-forms. 
Only for ${\rm E}_{8(8)}$ this symmetry comes down to the vector fields and becomes 
an integral part of the algebra of generalized diffeomorphisms.

In this paper, we construct the supersymmetric completion of the E$_{8(8)}$ exceptional field theory.
The theory is defined on a (3+248)-dimensional generalized spacetime. In addition to the usual dependency in spacetime (`external') coordinates $x^{\mu}$, $\mu=0, 1, 2$, all fields and gauge parameters formally depend also on extended coordinates $Y^\cM$, $\cM=1, \dots ,248$, transforming in the adjoint representation of $\rm{E}_{8(8)}$. As usual, not all of these internal coordinates are physical. This is taken care of by the E$_{8(8)}$ covariant section constraints, 
\bea
\label{section}
\eta^{\cM \cN}\,\partial_\cM \otimes {\partial}_\cN\equiv 0\;, \quad 
f^{\cM\cN\cK} \, \partial_\cM \otimes \partial_\cN\equiv0\;, \quad 
{(\P_{3875})_{\cM\cN}}^{\cK\cL}\partial_\cK \otimes {\partial}_\cL &\equiv&0\;,
\eea
where $\eta^{\cM\cN}$ and $f^{\cM\cN\cK}$ are respectively the Cartan-Killing form and the structure constants of E$_{8(8)}$ 
(see appendix~\ref{app:conventions} for more details  on the conventions used throughout this paper),
and
$\P_{\mathbf{3875}}$ is the projector onto the irreducible  representation $\mathbf{3875}$ in the tensor product of two adjoint representation
\bea 
\mathbf{248}\otimes \mathbf{248} &=& 
\mathbf{1}\oplus\mathbf{248}\oplus\mathbf{3875}\oplus\mathbf{27000}\oplus\mathbf{30380} \;,
\label{adjadj}
\eea
explicitly given by
\bea
(\P_{\mathbf{3875}})_{\cN\cL}{}^{\cM\cK}=\frac17\,\delta_{(\cN}{}^\cM\,\delta_{\cL)}{}^\cK
-\frac{1}{56}\,\eta^{\cM\cK}\,\eta_{\cN\cL}-\frac{1}{14}f^\cP{}_\cN{}^{(\cM}f_{\cP\cL}{}^{\cK)}
\;.
\eea 
The bosonic sector of the theory combines an external three-dimensional metric $g_{\mu\nu}$ 
(or dreibein $e_\mu{}^a$), an internal frame field ${\cal V}_{\cal M}{}^{\underline{{\cal K}}}$, parametrizing the coset
space ${\rm E}_{8(8)}/{\rm SO}(16)$, and gauge connections $A_\mu{}^{\cal M}$ and $B_{\mu\,{\cal M}}$
associated with generalized internal diffeomorphisms and constrained E$_{8(8)}$ rotations, respectively.
Fermions enter the theory as spinors under the ${\rm SO}(1,2)\times {\rm SO}(16)$ generalized Lorentz group
and transform as weighted scalars under generalized diffeomorphisms. 
Specifically, under ${\rm SO}(16)$, the gravitinos $\psi_\mu{}^I$ and fermions $\chi^{\dot{A}}$ 
transform in the fundamental vector ${\bf 16}$ and spinor ${\bf 128_c}$
representations, respectively. Their couplings require the introduction of spin connections
\bea
\begin{tabular}{c|ccc}
& ${\cal D}_\m$ & ${\cal D}_{{\cal M}}$ \\
\hline \\[-0.4cm]
${\rm SO}(1,2)$ & $\omega_\mu{}^{ab}$  & $\omega_{{\cal M}}{}^{ab}$ \\[0.15cm]
${\rm SO}(16)$ & ${\cal Q}_{\mu}{}^{IJ}$ & ${\cal Q}_{{\cal M}}{}^{IJ}$
\end{tabular}
\qquad
\;,
\label{spinspin}
\eea
in the external and internal coordinates, and for the two factors of the generalized Lorentz group, respectively.
In the external sector, 
the ${\rm SO}(1,2)$ connection $\omega_\m{}^{ab}$ is defined by the usual vanishing of external torsion according to
\bea
{\cal D}[A,\omega]_{[\mu} e_{\nu]}{}^a &=&0 \qquad \Longleftrightarrow\qquad \Gamma_{[\mu\nu]}{}^\rho ~=~0\;,
\eea
where (in contrast to standard geometry) derivatives are also covariantized w.r.t.\ internal generalized diffeomorphisms 
under which the dreibein $e_\mu{}^a$ transforms as a weighted scalar.
For the internal sector on the other hand, the generalized Christoffel connection is $\mathfrak{e}_{8(8)}$ valued
\bea
\Gamma_{{\cal MN}}{}^{{\cal K}}\equiv \Gamma_{{\cal M},{\cal L}}\,f{}^{{\cal LK}}{}_{{\cal N}}
\;,
\label{ChrisE8}
\eea
and the proper condition of vanishing torsion 
amounts to the projection condition~\cite{Cederwall:2015ica}
\bea
\left[\,
\Gamma_{{\cal M},{\cal N}}
\,\right]_{\,{\bf 1}\,\oplus\,{\bf 3875}} &=& 0\;, 
\eea
within the tensor product (\ref{adjadj}). 
As for the lower-rank exceptional groups~\cite{Coimbra:2011ky,Coimbra:2012af} this condition together
with a generalized vielbein postulate turns out to determine the internal ${\rm SO}(16)$
spin connection ${\cal Q}_{{\cal M}}{}^{IJ}$
up to contributions that drop out from all equations of motion and supersymmetry transformations
of the theory.
The off-diagonal blocks in (\ref{spinspin})
are determined by demanding that the algebra-valued currents
\bea
{\cal J}_{{\cal M}}{}^{ab} &\equiv& e^{a\m} \, {\cal D}[\omega]_{{\cal M}} e_\m{}^b\;,\qquad
{\cal J}_{\mu} ~\equiv~ 
{\cal V}^{-1}\,{\cal D}[{A},{B},{\cal Q}]_\mu {\cal V}~\in~\mathfrak{e}_{8(8)}
\;,
\eea
of the external and internal frame fields 
live in the orthogonal complement of the Lorentz algebra within ${\rm GL}(3)\times {\rm E}_{8(8)}$:
\bea
{\cal J}_{{\cal M}}\,\Big|_{\mathfrak{so}(1,2)} &=& 0 \;,\qquad
{\cal J}_{\mu}\,\Big|_{\mathfrak{so}(16)} ~=~ 0 \;.
\eea

The spin connections (\ref{spinspin}) are the central object for the description of fermionic couplings
and supersymmetry transformation rules in ${\rm E}_{8(8)} \times {\rm SO}(16)$-covariant form.
In this paper, we construct the unique supersymmetric completion of the bosonic
E$_{8(8)}$ exceptional field theory from~\cite{Hohm:2014fxa}.
Upon explicit solution of the section condition (\ref{section}), the resulting Lagrangian reduces to full 
$D=11$ supergravity and the IIB theory, respectively,
for appropriate reformulations of these theories, 
as pioneered in~\cite{Nicolai:1986jk,Melosch:1997wm,Koepsell:2000xg}.

The supersymmetric completion in particular underlines the role of the extra 
constrained gauge connection $B_{\mu\,{\cal M}}$ which joins the other fields in an
irreducible supermultiplet and whose variation contributes to the supersymmetry invariance
of the resulting action. 
Although its supersymmetry variation is given by some non-covariant expression, 
remarkably it turns out that
the following combination of variations
\bea
\Delta_\epsilon B_{\mu\,\cM} &\equiv& 
\delta_\epsilon B_{\mu\,\cM} -\Gamma_{\cM,\cN}\,\delta_\epsilon A_\mu{}^\cN
\;,
\label{DeltaB0}
\eea
takes a tensorial form
\bea
 \Delta_\epsilon B_{\mu\,\cM}
 &=&-2\left({\cal D}_{\cM}\bar{\epsilon}^I \psi_{\mu}{}^I-\bar{\epsilon}^I {\cal D}_{\cM} \psi_{\mu}{}^I\right)
+ie\,\varepsilon_{\mu\nu\rho}g^{\rho\sigma}{\cal D}_\cM(\bar{\epsilon}^I\gamma^{\nu}\psi_{\sigma}{}^I)
\;,
\eea
with supersymmetry parameter $\epsilon^I$.
The structure of the r.h.s.\ is such that it exhibits the full internal spin connection ${\cal Q}_{{\cal M}}{}^{IJ}$,
however its undetermined parts precisely cancel against the corresponding contributions from
the Christoffel connection $\Gamma_{{\cal M},{\cal N}}$
on the l.h.s., such that the net variation $\delta_\epsilon B_{\mu\,\cM}$ is uniquely determined
and compatible with the constraints this connection is subject to.

The paper is organized as follows.  After a brief review of the symmetry structure of
bosonic E$_{8(8)}$ exceptional field theory, we construct the necessary tools such as an internal fully covariant derivative and the ${\rm SO}(1,2)\otimes {\rm SO}(16)$ spin connections needed to 
describe the coupling to fermions and supersymmetry transformation rules.
We discuss their generalized curvatures whose components provide the building blocks for the bosonic field equations.
In section 3, we then analyze the algebra of supersymmetry transformations and
show that its closure into the bosonic symmetries of the theory entirely determines all the supersymmetry transformations. 
Finally, in section 4 we present the fermionic completion of the bosonic Lagrangian and prove its invariance under supersymmetry. 

%%%%%%%%%%%%%%%%%%%%%%%%%%%%%%%%%%%%%%%%%%%%%%
%%%%%%%%%%%%%%%%%%%%%%%%%%%%%%%%%%%%%%%%%%%%%%
\section{${\rm E}_{8(8)}\times {\rm SO}(16)$ exceptional geometry}\label{exgeo}
%%%%%%%%%%%%%%%%%%%%%%%%%%%%%%%%%%%%%%%%%%%%%%
%%%%%%%%%%%%%%%%%%%%%%%%%%%%%%%%%%%%%%%%%%%%%%

%%%%%%%%%%%%%%%%%%%%%%%%%%%%%%%%%%%%%%%%%%%%%%
\subsection{${\rm E}_{8(8)}$ generalized diffeomorphisms}
\label{subsec:gendiff}
%%%%%%%%%%%%%%%%%%%%%%%%%%%%%%%%%%%%%%%%%%%%%%

Let us start by a brief review of the
field content and symmetry structures of ${\rm E}_{8(8)}$ bosonic exceptional field theory.
For details, we refer to~\cite{Hohm:2014fxa}.
The field content is given by 
\bea
\{e_{\mu}{}^a\;,\, \cV_{\cM}{}^{\underline{{\cal K}}}\;,\, A_{\mu}{}^{\cM}\;,\,B_{\mu \, \cM} \}\;,
\label{fields_bos}
\eea
i.e.\ external and internal frame fields together with gauge connections $A_{\mu}{}^{\cM}$, $B_{\mu \, \cM}$\,.
The `dreibein' $e_{\mu}{}^a$ defines the external metric $g_{\mu\nu}=\eta_{ab}\,e_{\mu}{}^ae_{\nu}{}^b$.
The `248-bein' $\cV_{\cM}{}^{\underline{{\cal K}}}$ 
is the internal analogue of the dreibein and parametrises the coset space ${\rm E}_{8(8)}/{\rm SO}(16)$. Under SO(16), the collective index $\underline{{\cal K}}$ splits 
according to the decomposition of the algebra
\bea
\mathfrak{e}_{8(8)} &\longrightarrow&\mathfrak{so}(16) \oplus {\bf 128}_s
\;,
\label{e8so16}
\eea
into the adjoint and the spinor of SO(16), i.e.
\bea
\cV_{\cM}{}^{\underline{{\cal K}}}&=&\left\{\cV_{\cM}{}^{IJ},\cV_{\cM}{}^A\right\}\;,
\eea
satisfying $\cV_{\cM}{}^{IJ}=\cV_{\cM}{}^{[IJ]}$ with SO(16) vector indices $I, J = 1, \dots, 16$,
 and spinor indices $A, B = 1, \dots, 128$.\footnote{
 See appendix~\ref{app:conventions} for more details on the relevant group and algebra conventions.
 } 
In the same way the dreibein defines the external metric, the generalized vielbein defines the internal metric $\cM_{\cM\cN}$
\bea
\cM_{\cM\cN}&=&\cV_{\cM}{}^{\underline{{\cal K}}}\cV_{\cN}{}^{\underline{{\cal L}}}\,\delta_{\underline{\cal KL}}
~\equiv~ \cV_{\cM}{}^{A}\cV_{\cN}{}^{A}+\frac12\, \cV_{\cM}{}^{IJ}\cV_{\cN}{}^{IJ}\;,
\eea
in terms of which the bosonic theory can be formulated.
The inverse 248-bein then is given by
\bea
(\cV^{-1})
_{\underline{{\cal K}}}{}^{\cM}&=&\left\{\cV^{\cM}{}_B,-\cV^{\cM}{}_{I J}\right\}~\equiv~
\left\{ \eta^{\cM \cN}\cV_{\cN}{}^{B}, -\eta^{\cM \cN}\cV_{\cN}{}^{IJ}\right\} \;,
\eea
where
\bea
\cV_{\cM}{}^{A}\cV^{\cM}{}_B=\delta^{A}{}_B\;,\quad \cV_{\cM}{}^{IJ}\cV^{\cM}{}_{KL}=-2\,\delta^{IJ}_{KL}
\;.
\eea
Throughout, we raise and lower adjoint indices $\cM, \cN, \dots$, with the Cartan-Killing form $\eta_{\cM\cN}$.
Finally, the 248-bein is an ${\rm E}_{8(8)}$ group-valued matrix, which results in the 
standard decomposition of the Cartan form
\bea
(\cV^{-1})_{\underline{{\cal L}}}{}^{\cN}\partial_{\cM}\cV_{\cN}{}^{\underline{{\cal K}}}
&=&
\frac12\, q_{\cM}{}^{I J}\, (X_{I J}){}^{\underline{{\cal K}}}{}_{\underline{{\cal L}}}
+p_{\cM}{}^A \,(Y^A){}^{\underline{{\cal K}}}{}_{\underline{{\cal L}}}
\;,
\label{cartan}
\eea
where $X_{IJ}$ and $Y^A$ denote the compact and non-compact generators of ${\rm E}_{8(8)}$,
respectively. With the explicit expressions for the structure constants in the ${\rm SO}(16)$ basis from 
$\eqref{strcst}$, one finds the internal currents
\bea
q_{\cM}{}^{IJ}&=&\frac{1}{64}\,\Gamma^{IJ}_{BA}\mathcal{V}^{\cN}{}_B\partial_{\cM}\cV_{\cN}{}^A\;,\qquad
p_{\cM}{}^{B}~=~-\frac{1}{120}\,\Gamma^{IJ}_{BA}\cV^{\cN}{}_A\partial_{\cM}\cV_{\cN}{}^{IJ}\;,
\label{intcur}
\eea
which will be our building blocks for the internal spin connection and later the Ricci scalar.
This sums up the basic properties of the generalized vielbein.

The local symmetries of this exceptional field theory are generalized internal diffeomorphisms, 
constrained ${\rm E}_{8(8)}$ rotations, and external diffeomorphisms with 
respective parameters $\Lambda^\cM$, $\Sigma_\cM$, and $\xi^{\mu}$.
Let us first review the generalized internal diffeomorphisms.
The generalized Lie derivative acting on a vector $W^M$ of weight ${\lambda}_W$ is defined by
\bea
\L_{(\Lambda,\Sigma)}W^\cM= \Lambda^\cK{\partial}_\cK W^\cM -60\P^{\cM}{}_{\cN}{}^{\cK}{}_{\cL}\partial_\cK \Lambda^\cL W^\cN+\lambda_{W}{\partial}_\cN\Lambda^\cN W^\cM-{\Sigma}_{\cL}f^{\cL\cM}{}_\cN W^\cN
\;,
\label{GLie}
\eea
Here $\P^{\cM}{}_{\cN}{}^{\cK}{}_{\cL}$ projects onto the adjoint representation $\mathbf{248}$ and 
guarantees compatibility with the ${\rm E}_{8(8)}$ structure, c.f.\ the explicit expression~(\ref{Padj}).
The weight ${\lambda}_W$ of the various fields in the theory coincides with the three-dimensional Weyl 
weight of the fields, i.e.\ weight 2 and 0 for the external and internal metrics $g_{\mu\nu}$ and ${\cal M}_{\cM\cN}$,
respectively, and weights 1 and 0 for the gauge connections  $A_{\mu}{}^{\cM}$ and $B_{\mu \, \cM}$,
respectively. Fermions come with half-integer weight. This is summarized for all fields in Table~\ref{weights}.
\begin{table}[tb]
\begin{center}
\begin{tabular}{|c|c|c|c|c|c|c|}
\hline
Field & $e_{\mu}{}^a$ & $\cV_{\cM}{}^{\underline{K}}$ & $A_{\mu}{}^{\cM}$, $\Lambda^{\cM}$ & $B_{\mu \, \cM}$, $\Sigma_{\cM}$ & $\chi^{\dot{A}}$ & $\psi_{\mu}{}^I$, $\epsilon^I$ \\ \hline
Weight ($\lambda$)  & 1 & 0 & 1 & 0 & $-\frac12$ & $\frac12$ \\
\hline
\end{tabular}
\caption{Weights of all fields and gauge parameters 
under generalized diffeomorphisms.}
\label{weights}
\end{center}
\end{table}

Unlike the lower-rank ${\rm E}_{n(n)}$ cases with $n \leq 7$, the generalized Lie derivative (\ref{GLie}) depends on two parameters, $\Lambda^\cM$ and $\Sigma_\cM$, with the latter being subject to the section condition $\eqref{section}$,
i.e.
\bea
{(\P_{3875})_{\cM\cN}}^{\cK\cL}\,\Sigma_\cK \otimes \Sigma_\cL &\equiv&0~\equiv~
{(\P_{3875})_{\cM\cN}}^{\cK\cL}\,\Sigma_\cK \otimes {\partial}_\cL\;,\qquad
{\rm etc.}\;.
\label{sectionB}
\eea
This is needed together with the section constraints $\eqref{section}$
in order to ensure closure of the full symmetry algebra.
Schematically, we have an algebra
\bea
[\delta_{(\Lambda_1,\Sigma_1)},\delta_{(\Lambda_2,\Sigma_2)}]&=&\delta_{(\Lambda_{12},\Sigma_{12})}
\;,
\eea
with notably the gauge parameter $\Sigma_{12}$ given by
\bea
\Sigma_{12\,{\cal M}} \equiv
-2\,\Sigma_{[2\,{\cal M}} \partial_{{\cal N}} \Lambda_{1]}^{{\cal N}} 
+2\,\Lambda_{[2}^{{\cal N}} \partial_{{\cal N}} \Sigma_{1]\,{\cal M}} 
-2\, \Sigma_{[2}^{{\cal N}} \partial_{{\cal M}} \Lambda_{1]\,{\cal N}}
+f^{{\cal N}}{}_{{\cal KL}} \,\Lambda_{[2}^{{\cal K}}\,\partial_{{\cal M}}\partial_{{\cal N}}\Lambda_{1]}^{{\cal L}}
\;,
\eea 
confirming that the $\Lambda$ transformations do not close among themselves.

Before we describe the associated gauge connections and curvatures, let us make a small digression to discuss connections
and torsion compatible with the generalized diffeomorphisms (\ref{GLie}). 
For an algebra-valued connection
\bea
\Gamma_{{\cal MN}}{}^{{\cal K}}&=& 
\Gamma_{{\cal M},{\cal L}}\,f{}^{{\cal LK}}{}_{{\cal N}}
\;,
\label{Gamma}
\eea
the fact that pure $\Lambda$-transformations do not close into an algebra implies that the 
naive definition of torsion as
\bea
 \bar{\mathcal{T}}(\Lambda,W)^\cM&=&
 \bar{\cT}_{\cN\cK}{}^\cM\, \Lambda^\cN W^\cK~=~
 \L^{\nabla}_{(\Lambda,\Sigma)} W^\cM-\L_{(\Lambda,\Sigma)} W^\cM\;,
 \label{torsion_naive}
\eea
does no longer define a tensorial object. Here, $\L^{\nabla}$ refers to generalized Lie derivatives (\ref{GLie})
with partial derivatives replaces by covariant ones $\nabla=\partial-\Gamma$.
Following \cite{Cederwall:2015ica}, this suggests to rather define torsion as the part of the 
Christoffel connection that transforms covariantly under the 
generalized diffeomorphisms. 
With the transformation of (\ref{Gamma}) under (\ref{GLie}) given by
\bea
\delta_{(\Lambda,\Sigma)}{\Gamma}_{\cL,\cN}&=&
\delta^{\rm cov}_{(\Lambda,\Sigma)}\,{\Gamma}_{\cL,\cN}
+f_{\cQ\cN}{}^\cP{}\,{\partial}_\cL{\partial}_\cP{\Lambda}^\cQ+{\partial}_\cL\Sigma_\cN 
\label{christrans} \;,
\eea
projection onto its irreducible ${\rm E}_{8(8)}$ representations according to (\ref{adjadj})
shows that only its components in the ${\bf 1}\oplus{\bf 3875}$ transform as
tensors under (\ref{GLie}).
The proper definition of a torsionless connection thus corresponds to the condition
\bea
\left[\,
\Gamma_{{\cal M},{\cal N}}
\,\right]_{\,{\bf 1}\,\oplus\,{\bf 3875}} &=& 0\;,
\label{torsionless3875}
\eea
which can be made explicit with the form of the projector (\ref{3875proj}).
Let us note that such a torsionless connection gives rise to the identity
\bea
\L_{(\Lambda,\Sigma)}\, W^\cM&=& \L^{\nabla}_{(\Lambda,\tilde{\Sigma})}\, W^\cM
\;,
\label{SigSig}\\
&&{}
{\rm with}\qquad
\tilde\Sigma_{{\cal M}} \equiv \Sigma_{{\cal M}}-\Gamma_{{\cal M},{\cal N}}\,\Lambda^{{\cal N}}
\;.
\nonumber
\eea
With the r.h.s.\ of (\ref{SigSig}) manifestly covariant, this shows that the combination $\tilde\Sigma_{{\cal M}}$
behaves as a tensorial object under generalized diffeomorphisms. In this sense it may appear more natural
to parametrize generalized diffeomorphisms in terms of the parameters $(\Lambda, \tilde{\Sigma})$. 
The disadvantage of using $\tilde{\Sigma}$ w.r.t.\ the original formulation is the fact that the constraint (\ref{sectionB})
which $\Sigma_{\cM}$ has to satisfy, takes a much less transparent form when expressed in terms of $\tilde{\Sigma}$
since the connection $\Gamma_{\cM,\cN}$ in general will not be constrained in its first index
and will not even be fully determined by covariant constraints. 
For the description of generalized diffeomorphisms we thus have the choice between a description 
with covariant parameters $(\Lambda, \tilde{\Sigma})$ and a description in terms of parameters $(\Lambda, {\Sigma})$ 
in terms of which the constraints (\ref{sectionB}) are well defined and easily expressed. 
We will in general stick with the latter but observe that the existence of the covariant combination $\tilde{\Sigma}_\cM$
gives rise to some compact reformulations of the resulting formulas.\footnote{
The existence of the covariant combination $\tilde{\Sigma}_\cM$ may suggest to impose $\tilde{\Sigma}=0$
in order to reduce the number of independent gauge parameters~\cite{Cederwall:2015ica} while preserving closure of the algebra.
In view of the constraints (\ref{sectionB}), this is only possible in case the connection $\Gamma_{\cM,\cN}$ 
is identified with the Weitzenb\"ock connection 
$ \partial_\cM \cV_\cL{}^{\underline{\cP}}\,({\cal V}^{-1})_{\underline{P}}{}^\cK\,f_{\cN\cK}{}^\cL$ 
which itself is constrained in the first index.
We will in the following keep both gauge parameters $\Lambda^{{\cal M}}$ and $\Sigma_{{\cal M}}$
independent which seems important for the supersymmetric extension.}

The various terms of the bosonic action are constructed as
invariants under the generalized internal Lie derivatives (\ref{GLie}).
In the full theory, the gauge parameters $\Lambda^{\cM}$ and $\Sigma_{\cM}$ 
depend not only on the internal $Y^{{\cal M}}$ but also on the 
external $x^\mu$ coordinates. From the three-dimensional perspective, these
symmetries are implemented as (infinite-dimensional) gauge symmetries,
such that external derivatives are covariantized with gauge connections
$A_\mu{}^{{\cal M}}$, $B_{\mu\,{\cal M}}$
\bea
D_{\mu}=\partial_{\mu}-\L_{(A_{\mu},B_{\mu})}
\;.
\label{covD}
\eea
In accordance with (\ref{sectionB}), the connection $B_{\mu\,{\cal M}}$ is constrained 
to obey the same constraints as the gauge parameter $\Sigma_{{\cal M}}$.
The commutator of the covariant derivatives (\ref{covD}) closes into the field strengths
\bea
\label{commDD}
[D_\mu, D_\nu] &=& 
-\L_{({\cal F}_{\mu\nu},{\cal G}_{\mu\nu})}
\;,
\eea
with
\bea
   {\cal F}_{\mu\nu}{}^{\cM} 
   &= &2\,\partial_{[\mu}A_{\nu]}{}^{\cM}-2\,A_{[\mu}{}^{\cN}\partial_\cN A_{\nu]}{}^{\cM}
   +14\,(\mathbb{P}_{3875}){}^{\cM\cN}{}_{\cK\cL}
   A_{[\mu}{}^{\cK}\partial_\cN A_{\nu]}{}^{\cL}
   \nonumber\\
&&{}
   +\frac{1}{4}\,A_{[\mu}{}^{\cN}\partial^\cM A_{\nu]\cN}
   -\frac{1}{2}\,f^{\cM\cN}{}_{\cP}f^{\cP}{}_{\cK\cL}\,A_{[\mu}{}^{\cK}\partial_\cN A_{\nu]}{}^{\cL}
  ~+~\dots\;,
   \nonumber\\[2ex]
  {\cal G}_{\mu\nu \cM} &=& 2\,D_{[\mu}B_{\nu]\cM}
  -f^{\cN}{}_{\cK\cL}\,A_{[\mu}{}^{\cK}\partial_\cM\partial_\cN A_{\nu]}{}^{\cL}
  ~+~\dots
\;.
\label{FG}
\eea  
The ellipsis denote additional two-form terms 
required for the proper transformation behavior 
of the field strengths, c.f.\ (\ref{deltaFG}) below. 
As required for consistency, the section constraints (\ref{section}) 
ensure that all these terms drop from the commutators of covariant derivatives
where the field strengths are contracted with particular differential operators
according to (\ref{commDD}). Moreover, all the two-form terms drop out
from the bosonic Lagrangian. 

Under gauge transformations
\bea
\label{gaugeAB}
   \delta_{(\Lambda,\Sigma)} A_{\mu}{}^\cM &=& D_{\mu}\Lambda^\cM\;, \\
   \delta_{(\Lambda,\Sigma)} B_{\mu\, \cM} &=& D_{\mu}\Sigma_\cM-\Lambda^\cN\partial_\cM B_{\mu\, \cN}
   +f^{\cN}{}_{\cK\cL} \Lambda^\cK\partial_\cM\partial_\cN A_{\mu}{}^\cL\;,
\eea
(where just as the associated gauge connections, the parameters
$\Lambda^\cM$ and $\Sigma_\cM$ carry weight 1 and 0 under (\ref{GLie}), respectively,
c.f.\ Table~\ref{weights}),
the full field strengths (\ref{FG}) transform according to
\bea
\delta_{(\Lambda,\Sigma)}\,{\cal F}_{\mu\nu}{}^{\cM}&=&
\mathbb{L}_{(\Lambda,\Sigma)}\,{\cal F}_{\mu\nu}{}^{\cM}
\;,\nonumber\\
\delta_{(\Lambda,\Sigma)}\left(
{\cal G}_{\mu\nu\,\cM}-\Gamma_{\cM,\cN}\,{\cal F}_{\mu\nu}{}^\cN\right) 
&=&
\mathbb{L}_{(\Lambda,\Sigma)}
\left(
{\cal G}_{\mu\nu\,\cM}-\Gamma_{\cM,\cN}\,{\cal F}_{\mu\nu}{}^\cN\right) 
\;,
\label{deltaFG}
\eea
i.e.\ not the ${\cal G}_{\mu\nu\,\cM}$ but only the combination
$\tilde{{\cal G}}_{\mu\nu\,\cM}\equiv{\cal G}_{\mu\nu\,\cM}-\Gamma_{\cM,\cN}\,{\cal F}_{\mu\nu}{}^\cN$\,
behaves as a tensor under (\ref{GLie}).
This reflects the tensorial structure (\ref{SigSig}) of generalized diffeomorphisms.
Pushing this structure further ahead, we are led to introduce the general `covariant' variation
of the connection $B_{\mu\, \cM}$ as
\bea
\Delta B_{\mu\, \cM} &\equiv& \delta B_{\mu\, \cM}-\Gamma_{\cM,\cN}\,\delta A_\mu{}^\cN
\;,
\label{DeltaB}
\eea
c.f.~(\ref{DeltaB0}), in order to cast the gauge transformations (\ref{gaugeAB}) into
the more compact form
\bea
\label{gaugeABD}
   \delta_{(\Lambda,\Sigma)} A_{\mu}{}^\cM &=& D_{\mu}\Lambda^\cM\;, \nonumber\\
   \Delta_{(\Lambda,\Sigma)} B_{\mu\, \cM} &=& D_{\mu}\tilde\Sigma_\cM+\Lambda^\cN\,D_\mu\Gamma_{\cM,\cN}
   \;,
\eea
with $\tilde\Sigma_\cM$ from (\ref{SigSig}). This will turn out to be very useful in the following.

%%%%%%%%%%%%%%%%%%%%%%%%%%%%%%%%%%%%%%%%%%%%%%
\subsection{Section constraints}
%%%%%%%%%%%%%%%%%%%%%%%%%%%%%%%%%%%%%%%%%%%%%%

Since the section constraints (\ref{section}) play a central role in the construction
of the exceptional field theory, for the coupling of fermions it will be useful to spell out
the decomposition of these constraints under the subgroup ${\rm SO}(16)$ according to (\ref{e8so16}).
With the $\mathfrak{e}_{8(8)}$ representations of (\ref{section}) decomposing as
\bea
\mathbf{1\oplus 248\oplus3875}&\longrightarrow& \mathbf{1\oplus120\oplus128}_s \mathbf{\oplus135\oplus1820\oplus{1920}}_c\;,
\label{break_sectionso16}
\eea
the section constraints take the explicit form
\bea
{\cal M}^{\cM\cN} \,\partial_{\cM}\otimes \partial_{\cN} &=&
2\,{\cal V}^{\cM}{}_A{\cal V}^{\cN}{}_A\,\partial_{\cM}\otimes \partial_{\cN} 
\;,
\nonumber\\
{\cal V}^{\cM\,K[I}{\cal V}^{|\cN|\,J]K}\,\partial_{\cM}\otimes \partial_{\cN} &=&-\frac14 \Gamma^{IJ}_{CD} {\cal V}^{\cM}{}_C{\cal V}^{\cN}{}_D\,\partial_{\cM}\otimes \partial_{\cN}
\;,
\nonumber\\
\cV^{[\cM}{}_{IJ}\cV^{\cN]}{}_A\Gamma^{IJ}_{AB}\,\partial_{\cM}\otimes \partial_{\cN} &=&0
\nonumber\\
{\cal V}^{\cM\,K(I}{\cal V}^{|\cN|\,J)K}\,\partial_{\cM}\otimes \partial_{\cN} &=&-\frac{1}{16} \delta^{IJ} {\cal V}^{\cM}{}_{KL}{\cal V}^{\cN}{}_{KL}\,\partial_{\cM}\otimes \partial_{\cN}
\;,
\nonumber\\
{\cal V}^{\cM\,[IJ}{\cal V}^{|\cN|\,KL]}\,\partial_{\cM}\otimes \partial_{\cN} &=&
-\frac{1}{24} \Gamma^{IJKL}_{CD}\,{\cal V}^{\cM}{}_C{\cal V}^{\cN}{}_D\,\partial_{\cM}\otimes \partial_{\cN}\;,
\nonumber\\
\Gamma^J_{A\dot{A}}\cV^{(\cM}{\!}_{IJ}\cV^{\cN)}{\!\!}_A\;\partial_{\cM}\otimes \partial_{\cN}&=&-\frac{1}{16}\,(\Gamma^{MN}\Gamma^I)_{A\dot{A}}\;\cV^{(\cM}{\!}_{MN}\cV^{\cN)}{\!\!}_A\;\partial_{\cM}\otimes \partial_{\cN}\;,
\label{sectionso16}
\eea
which we will use in the following.
Following the above discussion, the same algebraic constraints hold for derivatives $\partial_{\cM}$
replaced by the gauge connection $B_{\mu\,\cM}$ or its gauge parameter $\Sigma_\cM$\,.

Let us recall from \cite{Hohm:2014fxa} that these section constraints allow for (at least) two inequivalent solutions
which break  E$_{8(8)}$ to GL$(8)$ or GL$(7)\times {\rm SL}(2)$,
and in which all fields depend on only eight or seven among the 248 internal coordinates $Y^\cM$, respectively.  
The resulting theory then coincides with the bosonic sector of $D=11$ and type IIB supergravity, respectively.

%%%%%%%%%%%%%%%%%%%%%%%%%%%%%%%%%%%%%%%%%%%
\subsection{Spin connections}
\label{subsec:spin}
%%%%%%%%%%%%%%%%%%%%%%%%%%%%%%%%%%%%%%%%%%%

Fermions enter the theory as spinors under the ${\rm SO}(1,2)\times {\rm SO}(16)$ generalized Lorentz group
and transform as weighted scalars under generalized diffeomorphisms. Their couplings thus require
four different blocks of the spin connection
\bea
\left\{
\begin{array}{ll}
 \omega_{\mu}& \quad  \omega_{\cM} \\
\cQ_{\mu}& \quad \cQ_{\cM}
\end{array}
\right\}
\label{qqq}
\eea
that ensure covariance of both external and internal derivatives under SO(1,2) and SO(16), respectively.
Via the generalized vielbein postulates 
\bea
0&\equiv&
\nabla_{\mu}e_{\nu}{}^a
~\equiv~
D_{\mu}e_{\nu}{}^a+\omega_{\mu}{}^{ab}e_{\nu}{}_b-\Gamma_{\mu\nu}{}^{\rho}e_{\rho}{}^a
\;,\nonumber\\
0 &\equiv&
\nabla_\cM\mathcal{V}_\cN{}^{\underline{K}}~\equiv~
\partial_\cM\mathcal{V}_\cN{}^{\underline{K}}-\frac12 \cQ_\cM{}^{IJ}(X_{IJ}){}^{\underline{K}}{}_{\underline{L}}\mathcal{V}_\cN{}^{\underline{L}}
-\Gamma_{\cM\cN}{}^\cP\,\mathcal{V}_\cP{}^{\underline{K}}
\;,
\label{GVP}
\eea
for the external and internal frame fields, the spin connections relate to 
the external and internal Christoffel connections
\bea
\left\{\, \Gamma_{\mu\nu}{}^{\rho},\, \Gamma_{\cM\cN}{}^{\cK}\, \right\}
\;.
\eea

Starting with the external sector, the SO(1,2) connection $\omega_{\mu}{}^{ab}$ 
is defined by the vanishing torsion condition of the external Christoffel connection
\bea
\Gamma_{[\mu\nu]}{}^\rho&=&0
\;. \label{exttorfree}
\eea
This leads to the standard expression for the spin connection in terms of the objects
of anholonomity $\Omega_{abc}\equiv 2\,e_{[a}{}^{\mu}e_{b]}{}^{\nu}\,D_{\mu}e_{\nu c}$,
where however derivatives are covariantized according to (\ref{covD}) with
the dreibein transforming as a scalar of weight 1 under (\ref{GLie}).
The external SO(16) connection on the other hand is defined by imposing that the external current
\bea
({\cal J}_{\mu})^{\underline{{\cal K}}}{}_{\underline{{\cal L}}}&\equiv&
(\cV^{-1})_{\underline{{\cal L}}}{}^{\cN}{\cal D}[A,{\cal Q}]_{\mu} \cV_{\cN}{}^{\underline{{\cal K}}}
\;,
\label{Jmu}
\eea
lives in the orthogonal complement of $\mathfrak{so}(16)$ within $\mathfrak{e}_{8(8)}$:
\bea
{\cal J}_{\mu}&\equiv&
{\cal P}_{\mu}{}^A \,Y^A
\;.
\label{P}
\eea
In analogy to (\ref{intcur}) this yields the explicit expressions
\bea
{\cal Q}_{\mu}{}^{IJ}&=&\frac{1}{64}\,\Gamma^{IJ}_{BA}\mathcal{V}^{\cN}{}_B \,D_\mu \cV_{\cN}{}^A\;,\qquad
{\cal P}_{\mu}{}^{B}~=~-\frac{1}{120}\,\Gamma^{IJ}_{BA}\cV^{\cN}{}_A \,D_\mu \cV_{\cN}{}^{IJ}\;,
\label{Qmu}
\eea
with covariant derivatives from (\ref{covD}).
According to their definition, the currents $\cP_{\mu}$ and $\cQ_{\mu}{}$ satisfy Maurer-Cartan integrability conditions
\bea
2\,\cD_{[\mu}\cP_{\nu]}{}^A&=&-\cF_{\mu\nu}{}^{\cM}p_{\cM A}+\cV_{\cP}{}^Af^{\cP \cM \cN}\partial_{\cM}\cF_{\mu\nu\, \cN}+\cG_{\mu\nu\,\cM}\cV^{\cM}{}_{A}\;,\label{intP}\\[2ex]
\cQ_{\mu\nu}{}^{I J}&\equiv&
2\,\partial_{[\mu}{\cal Q}_{\nu]}{}^{IJ} + 2\,{\cal Q}_{\mu}{}^{K[I} {\cal Q}_{\nu}{}^{J]K} 
\nonumber\\
&=&
-\cF_{\mu\nu}{}^{\cM}q_{\cM}{}^{IJ}+\cV_{\cP}{}^{IJ}f^{\cP \cM \cN}\partial_{\cM}\cF_{\mu\nu\,\cN}
+\cG_{\mu\nu \,\cM}\cV^{\cM}{}_{IJ}\nonumber\\
&&-\frac12\,\cP_{\mu}{}^A\cP_{\nu}{}^B\Gamma^{I J}_{AB}\label{intQ}
\eea
W.r.t.\ the integrability relations of $D=3$ supergravity~\cite{Marcus:1983hb},
these relations represent a deformation with additional terms in field strengths due to the introduction of the gauge fields $A_{\mu}{}^{\cM}$ and $B_{\mu\,\cM}$. We will see in the next section how these terms take a manifestly covariant form.
In the fermionic sector, 
the full external covariant derivatives acting on the ${\rm SO}(1,2)\times {\rm SO}(16)$ spinors of the theory are given by
\bea
\cD_{\mu}\psi^{I}&=&D_{\mu}\psi^{I}+\frac14\, \omega_{\mu}{}^{ab}\gamma_{ab}\,\psi^{I}+\cQ_{\mu}{}^{IJ}\,\psi^{J}\;,
\nonumber\\
\cD_{\mu}\chi^{\dot{A}}&=&D_{\mu}\chi^{\dot{A}}+\frac14\, \omega_{\mu}{}^{ab}\gamma_{ab}\,\chi^{\dot{A}}+\frac14\,\cQ_{\mu}{}^{IJ}\Gamma^{IJ}_{{\dot{A}}{\dot{B}}}\,\chi^{\dot{B}}\;,
\label{covFerm}
\eea
for spinors $\psi^I$ and $\chi^{\dot{A}}$ transforming in the ${\bf 16}$ and ${\bf 128}_c$ of SO(16), respectively.
Under generalized internal diffeomorphisms (\ref{GLie}), the spinors $\psi^I$ and $\chi^{\dot{A}}$ transform
as scalars of weight $1/2$ and $-1/2$, respectively, and the derivatives $D_\mu$ in (\ref{covFerm}) are
covariantized accordingly.

Now, let us turn to the internal sector. Similar to (\ref{P}) we derive the internal SO(1,2) spin connection by demanding
that the internal current
\bea
({\cal J}_\cM)^{ab} &\equiv& 
e^{b\mu}\,\cD[\omega]_{\cM}e_{\mu}{}^a\;,
\eea
lives in the orthogonal complement of $\mathfrak{so}(1,2)$ within $\mathfrak{gl}(3)$
\bea
({\cal J}_\cM)^{ab}\equiv \pi_{\cM}{}^{(ab)}\;.
\eea
Explicitly, this yields
\bea
\omega_{\cM}{}^{ab}=e^{\mu}{}^{[a}\partial_{\cM}e_{\mu}{}^{b]}\;.\label{spincon12}
\eea
In order to define the internal SO(16) connection, we recall that the proper condition of vanishing torsion
in the internal sector is given by setting to zero the tensorial part  (\ref{torsionless3875}) 
of the Christoffel connection $\Gamma_{\cM\cN}{}^{\cK}$.
Via (\ref{GVP}) this condition determines a large part of the SO(16) connection.
More precisely, the counting goes as follows~\cite{Cederwall:2015ica}: 
decomposition of (\ref{torsionless3875})
into SO(16) irreducible representations 
\bea
\mathbf{1\oplus3875}&\longrightarrow& \mathbf{1\oplus135\oplus1820\oplus{1920}}_c\;,
\label{torsionfreerep}
\eea
specifies the representation content of the vanishing torsion conditions.
On the other hand, the various components of the SO(16) connection $(\cQ_{\cM})^{IJ}$ live in
the SO(16) representations
\begin{equation}
\begin{aligned}
\cQ_{KL}{}^{IJ} \quad: \quad&\mathbf{120} \otimes \mathbf{120} =\mathbf{1} \oplus \mathbf{120} \oplus \mathbf{135} \oplus \mathbf{1820} \oplus\mathbf{5304} \oplus \mathbf{7020} \;,\\
\cQ_{A}{}^{IJ}\quad: \quad&\mathbf{120} \otimes \mathbf{128}_s = \mathbf{128}_s \oplus\mathbf{{1920}}_c \oplus\mathbf{13312}_s\;.
\end{aligned}
\label{Qfull}
\end{equation}
Comparison to (\ref{torsionfreerep}) exhibits which SO(16) components of $(\cQ_{\cM})^{IJ}$ are not fixed by
imposing vanishing torsion. For practical purposes, these undetermined parts 
${\bf 120}\oplus{\bf 128}_s\oplus{\bf 135}\oplus\mathbf{5304} \oplus \mathbf{7020}\oplus\mathbf{13312}_s$
do not pose a problem 
as they drop out of all physically relevant 
quantities such as the supersymmetry transformations, the Lagrangian etc., a property that all known 
supersymmetric exceptional field theories share.

Concretely, the four irreducible components (\ref{torsionfreerep})
of the torsion-free condition $\eqref{torsionless3875}$ 
take the form
\begin{equation}
\begin{aligned}
-\frac{1}{2}{\Gamma}_{IJ,IJ}+{\Gamma}_{A,A}=0\;,\\
-{\Gamma}_{M(I,J)M}-\frac{1}{16}\delta_{IJ}{\Gamma}_{MN,MN}=0\;,\\
{\Gamma}_{[IJ,KL]}+\frac{1}{24}\Gamma^{IJKL}_{AB}\,{\Gamma}_{A,B}=0\;,\\
\Gamma{}^J_{A\dot{A}}\left({\Gamma}_{IJ,A}+{\Gamma}_{A,IJ}\right)
+\frac{1}{16}\,(\Gamma{}^{MN}\Gamma^I)_{A\dot{A}}
\left({\Gamma}_{MN,A}+{\Gamma}_{A,MN}\right)=0\;.
\end{aligned}
\label{torsionfree}
\end{equation}
To explicitly solve these equations $\eqref{torsionfree}$, we use $\eqref{GVP}$, to 
express the internal Christoffel connection in terms of derivatives of the vielbein
\bea
{\Gamma}_{\cM, \cN}=\frac{1}{60 }f_{\cN}{}^{\cK \cP}\left( \cV_{\cP}{}^{A}\cD[{\cal Q}]_{\cM}\cV_{\cK}{}^A-\frac12 \cV_{\cP}{}^{IJ}\cD[{\cal Q}]_{\cM}\cV_{\cK}{}^{IJ} \right)
\;,
\label{chrisDV}
\eea
or, more explicitly
\bea
{\Gamma}_{\cM ,\cN}\cV^\cN{}_A=-p_{\cM, A}\,, \qquad {\Gamma}_{\cM, \cN}\cV^\cN{}_{IJ}=\cQ_{\cM}{}^{IJ}-q_{\cM}{}^{IJ}
\;,
\label{chrisexpl}
\eea
in terms of the Cartan form~(\ref{intcur}).
Then, combining these equations with $\eqref{torsionfree}$ translates conditions on the Christoffel connection into conditions on the spin connection. The solution for the SO(16) spin connection is then found to be
\bea
\cQ_{\cM}{}^{IJ}&=&{\cal V}_\cM{}^{A}\,\cQ_{A}{}^{IJ}-\frac12
{\cal V}_\cM{}^{KL}\,\cQ_{KL}{}^{IJ}
\;,
\eea
with
\bea
\cQ_{IJ}{}^{KL} &=& q_{IJ}{}^{KL}-\frac{1}{60}\,\delta^{KL}_{IJ}p_{A,A}+\frac{1}{14}\,\delta\!\oversym{^{I[K}_{\vphantom{M}}\,\Gamma_{AB}^{L]J}}p_{A,B}
\nonumber\\
&&+\frac{1}{4!}\,\Gamma^{IJKL}_{AB}\,p_{A,B}+\frac17\,\delta\!\oversym{^{I[K}_{\vphantom{M}}\,\cV_\cM{}_{\vphantom{M}}^{L]J}}\Gamma_{\cN}{}^{\cM\cN}+U_{IJ,KL}\;,\nonumber\\
\cQ_A{}^{IJ} &=& q_{A}{}^{IJ}+p_{IJ,A}-\frac{1}{56}\,\Gamma^{I K}_{AB}\,p_{K J,B}\,+\frac{1}{56}\,\Gamma^{J K}_{AB}\,p_{K I,B}\nonumber\\
&&+\frac{3}{364}\,\Gamma^{IJKL}_{AB}\,p_{KL,B}+\frac{1}{60}\,\cV^{\cM}{}_B\Gamma^{IJ}_{AB}\Gamma_{\cN\cM}{}^{\cN}\,+(R_{{13312}})_{A}{}^{IJ} \;,\label{spincon16}
\eea
c.f.~\cite{Cederwall:2015ica}, 
in terms of the Cartan forms~(\ref{intcur}), whose first indices we have `flattened' with the 248-bein 
${\cal V}_{\cM}{}^{\underline{\cK}}$.
The contributions $U_{IJ,KL}$, $(R_{{13312}})_{A}{}^{IJ}$ in (\ref{spincon16})
are constrained by
\bea
U_{IJ,KL}&=&U_{[IJ],[KL]}\;,\qquad
U_{[IJ,KL]}~=~ 0~=~  U_{IK,KJ}
\;,\nonumber\\
(R_{{13312}})_{A}{}^{IJ}&=&
(R_{{13312}})_{A}{}^{[IJ]}\;,\qquad
\Gamma^{I}_{A\dot{A}}\,(R_{{13312}})_{A}{}^{IJ}~=~ 0\;,
\eea
and not determined by the vanishing torsion condition,
in accordance with (\ref{Qfull}). The undetermined parts in the ${\bf 120}\oplus{\bf 128}_s$ in (\ref{spincon16})
have been expressed via the trace $\Gamma_{\cN\cM}{}^{\cN}$ of the
Christoffel connection. The latter can be fixed by imposing as an additional condition that 
the determinant of the external vielbein $e\equiv{\rm det}\,e_\mu{}^a$
be covariantly constant
\bea
\nabla_{\cM}e~\equiv~\partial_{\cM}e-\frac32 \,\Gamma_{\cN \cM}{}^{\cN} e~\equiv~ 0\;, \qquad \Longrightarrow \qquad
\Gamma_{\cN \cM}{}^{\cN}~=~\frac23 \, e^{-1}\partial_{\cM}e\;.
\eea
To summarize, the full internal covariant derivative act on an ${\rm E}_{8(8)}\times{\rm SO}(16)$ tensor $X_{\cM}{}^I$ 
of weight $\lambda_X$ as
\bea
\nabla_\cM X_{\cN}{}^I &\equiv&
\partial_\cM X_{\cN}{}^I + {\cal Q}_{\cM}{}^{IJ} X_\cN{}^J-\Gamma_{\cM\cN}{}^\cK X_{\cK}{}^I-\frac12 \lambda_X \Gamma_{\cK\cM}{}^\cK X_{\cN}{}^I\;,
\label{nablaM}
\eea
with the connections defined by (\ref{spincon16}) and (\ref{chrisexpl}), respectively.
This covariant derivative transforms as a 
generalized tensor of weight $\lambda=\lambda_X-1$ under generalized diffeomorphisms.
In particular, for the spinor fields of the theory, the covariant internal derivatives take the form
\bea
\nabla_\cM \psi_{\mu}^I &\equiv&
\partial_\cM \psi_{\mu}^I + {\cal Q}_{\cM}{}^{IJ} \psi_{\mu}^I 
+\frac14\,\omega_{\cM}{}^{ab}\,\gamma_{ab}\,\psi_{\mu}^I 
-\frac14 \, \Gamma_{\cK\cM}{}^\cK \,\psi_{\mu}^I \;,\nonumber\\
\nabla_\cM \chi^{\dot{A}} &\equiv&
\partial_\cM \chi^{\dot{A}}  +\frac14\, {\cal Q}_{\cM}{}^{IJ} \,\Gamma^{IJ}_{\dot{A}\dot{B}}\,\chi^{\dot{B}}  
+\frac14\,\omega_{\cM}{}^{ab}\,\gamma_{ab}\,\chi^{\dot{A}} 
+\frac14  \, \Gamma_{\cK\cM}{}^\cK \,\chi^{\dot{A}}  \;.
\label{nablaMFerm}
\eea

We conclude this section with a collection of the different covariant derivatives we have used
and will use throughout this paper:
\begin{alignat}{2}
D_{\mu}&=D[A]_{\mu} \;,\qquad \qquad &&\nonumber\\
\cD_{\mu}&=\cD[A,\omega, \cQ]_{\mu}\;, \qquad \qquad && \cD_{\cM}=\cD[\omega,\cQ]_{\cM}\;,\nonumber\\
\nabla_{\mu}&= \nabla[A,\omega,\cQ,\Gamma]_{\mu} \;, \qquad \qquad && \nabla_{\cM}=\nabla[\omega,\cQ,\Gamma]_{\cM}
\;,
\end{alignat}
where $A_{\mu}{}^{\cM}$ is the gauge field associated with generalized diffeomorphisms symmetry and the four blocks of the spin connection $\omega_{\mu},\cQ_{\mu},\omega_{\cM},\cQ_{\cM}$ defined in $\eqref{exttorfree}, \eqref{Qmu}, \eqref{spincon12}, \eqref{spincon16}$, respectively.

\subsection{Curvatures}

Having defined the various components of the spin connection (\ref{qqq}), we can now discuss
their curvatures which will be the building blocks for the bosonic Lagrangian and field equations.
Moreover, we will require a number of identities for the commutators of covariant derivatives in order
to prove the invariance of the full Lagrangian under supersymmetry.

Let us start with the commutator of two external covariant derivatives on an ${\rm SO}(1,2)\times {\rm SO}(16)$ spinor 
$\epsilon^I$ which is obtained straightforwardly from (\ref{covFerm})
\bea
\left[\cD_{\mu},\cD_{\nu}\right]\epsilon^I&=&
-\cF_{\mu\nu}{}^\cM\partial_{\cM}\epsilon^I-\frac12\partial_{\cM}\cF_{\mu\nu}{}^\cM \epsilon^I+\cQ_{\mu\nu}{}^{I J}\epsilon^J+\frac14\cR_{\mu\nu}{}^{ab}\gamma_{ab}\epsilon^I\;,
\label{comderext}
\eea
with the field strength of the gauge field $A_{\mu}{}^\cM$ introduced in $\eqref{FG}$, 
the usual external Riemann curvature defined by 
\bea
\cR_{\mu\nu}{}^{ab}=2 D_{[\mu}\omega_{\nu]}{}^{ab}+2\omega_{[\mu}{}^{ac}\,\omega_{\nu]\,c}{}^b\;,
\label{extRiemann}
\eea 
(with covariant derivatives (\ref{covD})), 
and its analogue $\cQ_{\mu\nu}{}^{I J}$ from (\ref{intQ}) for the SO$(16)$ external spin connection.
As the commutator of two external covariant derivatives,
the left-hand side of $\eqref{comderext}$ is covariant
whereas this is not manifest from the r.h.s..
Embedding the internal derivatives on the r.h.s.\ into full covariant derivatives (\ref{nablaMFerm}),
the commutator can be rewritten as
\bea
\left[\cD_{\mu},\cD_{\nu}\right]\epsilon^I&=&
-\cF_{\mu\nu}{}^\cM\nabla_{\cM}\epsilon^I
-\frac12\nabla_{\cM}\cF_{\mu\nu}{}^\cM \epsilon^I\epsilon^J+\frac14\widehat{\cR}_{\mu\nu}{}^{ab}\gamma_{ab}\epsilon^I
\nonumber\\
&&{}
+\cQ_{\mu\nu}{}^{I J}+\cF_{\mu\nu}{}^{\cM}\cQ_{\cM}{}^{IJ}\epsilon^J
\label{commdd1}
\eea
with the improved Riemann tensor 
$\widehat{\cR}_{\mu\nu}{}^{ab}\equiv\cR_{\mu\nu}{}^{ab}+\omega_{\cM}{}^{ab}\cF_{\mu\nu}{}^{\cM}$.
The latter is covariant under local SO(1,2) Lorentz transformations,
shows up in the gravitational field equations and
whose contraction in particular gives rise to the improved Ricci scalar
\bea
\widehat{R}&=& e_a{}^\mu e_b{}^\nu\,\widehat{\cR}_{\mu\nu}{}^{ab}
\;,
\label{REH}
\eea
that is part of the bosonic action.
With the first line of (\ref{commdd1}) now manifestly covariant, the second line can be rewritten upon
using the explicit expression (\ref{intQ}) for $\cQ_{\mu\nu}{}^{I J}$ such that the
commutator takes the manifestly covariant form
\bea
\left[\cD_{\mu},\cD_{\nu}\right]\epsilon^I&=&
\frac12\,\cP_{\mu}{}^A\cP_{\nu}{}^B\Gamma^{I J}_{AB}\,\epsilon^J+\frac14\widehat{\cR}_{\mu\nu}{}^{ab}\gamma_{ab}\,\epsilon^I+\cV_{\cP}{}^{I J}f^{\cP \cM \cN}\,\nabla_{\cM}\cF_{\mu\nu\,\cN}\,\epsilon^J\nonumber\\
&&+\widetilde{\cG}_{\mu\nu \,\cM}\cV^{\cM}{}_{IJ}\,\epsilon^J-\cF_{\mu\nu}{}^\cM\nabla_{\cM}\,\epsilon^I-\frac12\nabla_{\cM}\cF_{\mu\nu}{}^\cM\,\epsilon^I
\label{comderextcov}\;,
\eea
with the tensorial combination of field strengths $\widetilde{\cG}_{\mu\nu \,\cM}$ from (\ref{deltaFG}).
Similarly, one may rewrite the second integrability relation $\eqref{intP}$ into the manifestly covariant form
\bea
2\,\cD_{[\mu}\cP_{\nu]}{}^A&=&\cV_{\cP}{}^{A}f^{\cP \cM \cN}\nabla_{\cM}\cF_{\mu\nu\, \cN}+\widetilde{\cG}_{\mu\nu\,\cM}\cV^{\cM}{}_{A}\label{intP1}\;.
\eea

We now turn to the mixed curvature, arising from the commutators of one external and one internal covariant derivatives. We will only be interested in those projections of this commutator, in which the undetermined part of the SO(16) connection drops out. Fortunately, they are the projections relevant to prove the invariance of the Lagrangian under supersymmetry. Evaluating different projections of such a commutator on an ${\rm SO}(1,2)\times {\rm SO}(16)$ spinor 
$\epsilon^I$, we obtain the relations
\bea
\cV^{\cM}{}_{A}\Gamma^I_{A\dot{A}}
\,
\left[\nabla_{\cM}, {\cal D}_\mu \right]{\epsilon}^I
&=&
\frac14\,\cV^{\cM}{}_{A}\Gamma^I_{A\dot{A}}
\,
{\cal R}_{\cM \mu}{}^{ab}\,\gamma_{ab} {\epsilon}^I
\nonumber\\
&&{}
-\frac34\,\Gamma^{I}_{A\dot{A}} {\cal V}^\cM{}_{IJ} \nabla_\cM {\cal P}_\mu{}^A \,\epsilon^J
+\frac18\,\Gamma^{IJK}_{A\dot{A}} {\cal V}^\cM{}_{IJ} \nabla_\cM {\cal P}_\mu{}^A \,\epsilon^K\;,
\nonumber\\[2ex]
\cV^{\cM}{}_{IJ}
\,
\left[\nabla_{\cM}, {\cal D}_\mu \right]{\epsilon}^J
&=&
\frac14\,\cV^{\cM}{}_{IJ}
\,
{\cal R}_{\cM \mu}{}^{ab}\,\gamma_{ab} {\epsilon}^J
\nonumber\\
&&{}
-\frac18\,{\cal V}^\cM{}_A \nabla_\cM {\cal P}_\mu{}^A \,\epsilon^I
-\frac14\,\Gamma^{IJ}_{AB} {\cal V}^\cM{}_A \nabla_\cM {\cal P}_\mu{}^B \,\epsilon^J\;,
\eea
where  the mixed curvature tensor is defined by
\bea
\cR_{\cM\mu}{}^{\nu\rho}= e_a{}^{\nu}e_b{}^{\rho}\left(\partial_{\cM}\omega_{\mu}{}^{ab}-\cD_{\mu}\omega_{\cM}{}^{ab}\right)
=\left(\partial_{\cM}\Gamma_{\mu\sigma}{}^{[\nu}\right)g^{\rho]\sigma}\;.
\eea
One can show it constitutes a tensor under generalized diffeomorphisms (\ref{GLie}),
and satisfies a Bianchi identity
\bea
\cR_{\cM[\mu\,\nu\rho]}\equiv 0\;.
\eea
Its contraction to a `mixed Ricci tensor' yields the following current
\bea
{{\cal R}}_{\cM\nu}{}^{\mu\nu} &=&
-\frac12\, \widehat{J}^\mu{}_\cM
~\equiv~
 e_{a}{}^\mu e_{b}{}^\nu \left(
\partial_M \omega_{\nu}{}^{ab} 
-  {\cal D}_{\nu} \left(   e^{\rho[a} \partial_M e_{\rho}{}^{b]} \right) \right)
\;,
\label{BIR2}
\eea
which is related to the improved Ricci scalar (\ref{REH}) by variation w.r.t.\ the vector fields
\bea
\delta_A\widehat{R}&=& \widehat{J}^{\mu}{}_\cM\, \delta A_{\mu}{}^\cM
~+~\nabla_\cM {\cal J}_A^\cM +\cD_{\mu}\cI_{A}^{\mu}
\;,
\eea
up to a boundary currents ${\cal J}_A^\cM$, $\cI_{A}^{\mu}$ of respective weights $\lambda_{{\cal J}_A}=-1$,
$\lambda_{\cI_{A}}=-2$\,,
that do not contribute under the integral.

Finally, for the internal curvature, we are again interested in specific projections 
of two internal covariant derivative in which the undetermined part of the connection drops out. 
The pertinent projection for the definition of an internal curvature scalar ${\cal R}$
in the ${\rm E}_{8(8)}\times {\rm SO}(16)$ exceptional geometry is given by
\bea
&&\left(16\,\cV^{\cM}{}_{KI}\cV^{\cN}{}_{JK}
+2\,{\cal V}^{\cM}{\!}_A{\cal V}^{\cN}{\!}_A\,\delta_{I J}
+2\,\Gamma^{IJ}_{AB}\cV^{\cM}{}_{A}\cV^{\cN}{}_{B}
\right)
\nabla_{\cM}\nabla_{\cN}\,\epsilon^J ~=
\nonumber\\
&& \qquad\qquad\qquad\qquad\qquad\qquad\qquad
=~
-\frac{1}{8} \,\cR\, \epsilon^I
+\cV^{\cM}{}_{KI}\cV^{\cN}{}_{JK} \,\cR_{\cM \cN}{}^{a b}\gamma_{a b}\epsilon^J\;.
\qquad\qquad
\label{comintder}
\eea
On the l.h.s.\ the double derivative terms vanish by means of the 
section constraints (\ref{sectionso16}), while a straightforward computation
shows that also all linear derivative terms $\partial_\cM\epsilon^I$ cancel.
The curvature of the internal spin connection on the r.h.s.\ is defined in analogy to (\ref{extRiemann}) and
computed to be
\bea
\cR_{\cM \cN}{}^{ab}&=&2\,\partial_{[\cM}\omega_{\cN]}{}^{ab}+2\omega_{[\cM}{}^{ac}\,\omega_{\cN] \,c}{}^b
\nonumber\\
&=& 
  - \frac{1}{2} \, e^{\mu[a}e^{b]\nu}g^{\sigma \tau} \nabla_{\cM} g_{\mu \sigma} \nabla_{\cN} g_{\nu \tau} 
\;.
\label{RMNab}
\eea
Upon using the expressions for the SO(16) spin connection (\ref{spincon16}),
the internal curvature scalar $\cR$ in (\ref{comintder}) can be calculated explicitly in terms of the 
Cartan forms (\ref{intcur}) and the derivative of the external vielbein determinant $e$ as
\bea
\cR&=&
-\frac23\,{\cal M}^{\cM\cN} e^{-2} \partial_\cM e\,\partial_\cN e
+\frac43\,{\cal M}^{\cM\cN}  e^{-1} \partial_\cM \partial_\cN e
+\frac43\,
{\cal V}^{(\cM}{}_A {\cal V}^{\cN)}{}_{IJ}\,
\Gamma^{IJ}_{AB}\,p_{\cM}{}^B \,e^{-1}\partial_{\cN}e
\nonumber\\
&&{} 
+{\cal V}^{\cM}{}_A{\cal V}^{\cN}{}_{IJ}\,\Gamma^{IJ}_{AB} 
\left(\partial_{(\cM} p_{\cN)}{}^B+\frac14 \Gamma^{IJ}_{BC} \,q_{(\cM}{}^{IJ}p_{\cN)}{}^C
\right)
+{\cal \cM}^{\cM\cN} \,p_\cM{}^A\,p_\cN{}^A
\nonumber\\
&&{}+ 2\, {\cal V}^{\cM}{}_A{\cal V}^{\cN}{}_B\,p_\cM{}^B p_\cN{}^A
-\frac18 {\cal V}^{\cM}{}_{IJ} {\cal V}^{\cN}{}_{KL} \,
\left(\Gamma^{IJ} \Gamma^{KL}\right)_{AB} \,p_\cM{}^A p_\cN{}^B
\nonumber\\
&&{}
+\frac14\, {\cal V}^{\cM}{}_A{\cal V}^{\cN}{}_B\,
\Gamma^{IJ}_{AC}\Gamma^{IJ}_{BD}\,p_\cM{}^C p_\cN{}^D 
\;.
\label{curvsca}
\eea
By construction it transforms as a scalar (of weight $\lambda_{{\cal R}}=-2$) under generalized
diffeomorphisms (\ref{GLie}).
Its dependence on the external metric is such that
\bea
\delta (e{\cal R}) &=& (\delta e) \,{\cal R}~+\mbox{total derivatives}\;.
\label{varR2}
\eea

The other relevant projection of two internal derivatives on
a spinor is given by
\bea
&&\left(12\, \cV^{\cM}{}_{A}\cV^{\cN}{}_{I J}\,\Gamma^{I}_{A\dot{A}}
+\left(\Gamma^{I J K}_{A\dot{A}}
+2\Gamma^{I}_{A\dot{A}}\delta^{J K}\right)
\cV^{\cM}{}_{I K}\cV^{\cN}{}_{A}\right)\nabla_{\cM}\nabla_{\cN}\,\epsilon^J
~=
\nonumber\\
&& \qquad\qquad
=~
\frac18\,\Gamma^{I}_{A\dot{A}}\cR_{A}\,\epsilon^I
+\frac{1}{16}\, \cV^{\cM}{}_{I J}\cV^{\cN}{}_{A}
\left(
\Gamma^{I J K}_{A \dot{A}}-14\,\delta^{JK}\Gamma^{I}_{A \dot{A}}
\right)
\cR_{\cM \cN}{}^{a b}
\gamma_{a b}\,\epsilon^K
\;,\qquad
\label{comintder2}
\eea
where again all double derivatives on the l.h.s.\ vanish due to the section constraints.
The generalized curvature $\cR_A$ on the r.h.s.\ plays the analogue
of a Ricci tensor in this geometry and is
most conveniently defined by variation of the curvature scalar ${\cal R}$ w.r.t.\
to a non-compact local $\mathfrak{e}_{8(8)}$ transformation of the internal frame field, i.e.
\bea
 \delta_\Sigma \cR &\equiv& \Sigma^A(Y) \,\cR_A
~+~\nabla_\cM {\cal J}_\Sigma^\cM 
\;,\qquad
\mbox{under}\quad
\delta_\Sigma {\cal V} ~=~ {\cal V}\,Y^A\Sigma^A(Y)
\;.
\eea
up to a boundary current ${\cal J}_\Sigma^\cM$ of weight $\lambda_{{\cal J}_\Sigma}=-1$\,.
It can be explicitly given in terms of the 
Cartan forms (\ref{intcur}) as
\bea
\cR_A &=& - \frac{2}{3}\, {\Gamma}^{I M}_{ A B} \cV^{\cM}{}_{I M} \cV^{\cN}{}_{B} \partial_{\cM}e\,  \partial_\cN e\,  e^{-2}+\frac{1}{4}\, {\Gamma}^{I M}_{A B} {\Gamma}^{I M N P}_{ C D} \cV^{\cM}{}_{N P} \cV^{\cN}{}_{B} {p}_{\mathcal{M}}{}^C {p}_{\mathcal{N}}{}^D
\nonumber\\ 
&& 
- \, {\Gamma}^{I M}_{ A B} {\Gamma}^{I N}_{ C D} \cV^{\cM}{}_{M N} \cV^{\cN}{}_{B} {p}_{\mathcal{M}}{}^C {p}_{\mathcal{N}}{}^D - \frac{3}{2}\, {\Gamma}^{I M}_{A B}\cV^{\cM}{}_{I M} \cV^{\cN}{}_{B} {p}_{\mathcal{M}}{}^C {p}_{\mathcal{N}}{}^C\nonumber\\  
&&- 2\, {\Gamma}^{I M}_{ C B}\cV^{\cM}{}_{I M} \cV^{\cN}{}_{C} {p}_{\mathcal{M}}{}^A {p}_{\mathcal{N}}{}^B 
+ \frac{23}{16}\, {\Gamma}^{I M}_{ A B} {\Gamma}^{I N}_{ C D} \cV^{\cM}{}_{M N} \cV^{\cN}_{C} {p}_{\mathcal{M}}{}^D {p}_{\mathcal{N}}{}^B\nonumber\\
&& + {\Gamma}^{I M}_{ C B} \cV^{\cM}{}_{I M} \cV^{\cN}{}_{A} {p}_{\mathcal{M}}{}^C {p}_{\mathcal{N}}{}^B
+2\, {\Gamma}^{I M}_{ A B} \cV^{\cM}{}_{I M} \cV^{\cN}{}_{C} {p}_{\mathcal{M}}{}^B {p}_{\mathcal{N}}{}^C\nonumber\\
 &&+\Big(- 4\,  \cV^{\cN}{}_{A} \cV^{\cM}{}_{B}
 - 3\, \delta_{AB}\,  \cV^{\cM}{}_{C} \cV^{\cN}{}_{C}
- \frac{1}{4}\, {\Gamma}^{I M N P}_{A B}\,  \cV^{\cM}{}_{I M} \cV^{\cN}{}_{N P} \nonumber\\
&&\qquad + \frac{1}{2}\, {\Gamma}^{I M}_{ A C} {\Gamma}^{I M}_{B D}\,  \cV^{\cN}{}_{C} \cV^{\cM}{}_{D}\Big)\left(\partial_{(\cM} p_{\cN)}{}^B+\frac14 \Gamma^{IJ}_{BC} \,q_{(\cM}{}^{IJ}p_{\cN)}{}^C
\right).
\eea
This expression above is given in compact form, 
after simplification by various Fierz-like identities, 
some of which are collected in appendix~\ref{gamid}.

%%%%%%%%%%%%%%%%%%%%%%%%%%%%%%%%%%%%%%%%%
%%%%%%%%%%%%%%%%%%%%%%%%%%%%%%%%%%%%%%%%%
\section{Supersymmetry algebra}
%%%%%%%%%%%%%%%%%%%%%%%%%%%%%%%%%%%%%%%%%
%%%%%%%%%%%%%%%%%%%%%%%%%%%%%%%%%%%%%%%%%

In this section we establish the supersymmetry transformation of the various fields and verify that
the supersymmetry algebra closes. Before discussing supersymmetry, we briefly review the
bosonic symmetries of E$_{8(8)}$ exceptional field theory, since these are the transformations we
are going to recover in the commutator of two supersymmetry transformations.

\subsection{Bosonic symmetries of E$_{8(8)}$ exceptional field theory}

In section~\ref{subsec:gendiff} we have extensively discussed the structure of internal generalized
Lie derivatives which depend on two parameters $\Lambda^\cM$ and $\Sigma_\cM$ with associated
gauge connections $\cA_\mu{}^\cM$ and $B_{\mu\,\cM}$\,.
A closer analysis \cite{Hohm:2014fxa} shows that these gauge connections come with additional
shift symmetries which take the form
\bea
\delta_\Xi A_{\mu}{}^{\cM}&=&\partial_{\cK}\Xi_{\mu}{}_{3875}{}^{\cM \cK}
+\eta^{\cM \cN}\Xi_{\mu\, \cN}
+f^{\cM \cN}{}_{\cK}\Xi_{\mu\, \cN}{}^{\cK}\;,\nonumber\\
\delta_\Xi B_{\mu \,\cM}&=&\partial_{\cM}\Xi_{\mu\,\cN}{}^{ \cN}+\partial_{\cN}\Xi_{\mu\,\cM}{}^{ \cN}
\;.
\label{shiftAB}
\eea
Here, the symmetry parameter $\Xi_{\mu}{}_{3875}{}^{\cM \cN}$ lives in the projection of the
two adjoint indices $\cM\cN$ onto the ${\bf 3875}$ representation, explicitly realized by~(\ref{3875proj}). 
The parameter
$\Xi_{\mu\, \cN}$ is constrained in the same way as the fields 
$B_{\mu\,\cM}$ and $\Sigma_{\cM}$, c.f.~(\ref{sectionB}).
Similarly, the parameter $\Xi_{\mu\, \cN}{}^{\cK}$ is constrained as (\ref{sectionB}) in its first 
internal index $\cN$.
It is straightforward to check that the shift symmetries (\ref{shiftAB}) 
leave the covariant derivatives (\ref{covD})
invariant. More precisely, they correspond to the tensor gauge transformations associated to
the two-form gauge fields that complete the vector field strengths ${\cal F}_{\mu\nu}{}^{\cM}$
and ${\cal G}_{\mu\nu\,\cM}$ into fully covariant objects, but drop out from the Lagrangian
of the theory.

Apart from the internal gauge symmetries, the full set of bosonic symmetries also includes a 
covariantized version of the (2+1)-external diffeomorphism
with the parameter $\xi^{\mu}$ depending on both set of coordinates $\{x^{\mu},Y^\cM\}$.
On the bosonic fields these act as\footnote{
W.r.t.\ the form of these transformations given in \cite{Hohm:2014fxa},
we have expressed the current bosonic current $j^{\rho \cM}$ by
the coset current $\cP^{\rho \cM}$, see~(\ref{JinP}) below, and 
furthermore changed the vector transformations
by a shift transformation (\ref{shiftAB}) with parameter
$\Xi_{\mu\,\cM}=-g_{\mu\nu}\partial_{\cM}\xi^\nu$, in order
to obtain a more compact presentation of the external diffeomorphisms. 
Also some signs differ from the formulas in \cite{Hohm:2014fxa}
due to the fact that  in this paper we use mostly minus signature $(+--)$
for the external metric.
}
\bea
\delta_{\xi}e_{\mu}{}^a&=&\xi^{\nu}D_{\nu}e_{\mu}{}^a+D_{\mu}\xi^\nu e_{\nu}{}^a\;,\nonumber\\
\delta_{\xi}\cM_{\cM\cN}&=&\xi^{\nu}D_{\nu}\cM_{\cM\cN}\;,\nonumber\\
 \delta_{\xi} A_{\mu}{}^{\cM} &=& 
  -2\,{\cal V}^{\cM A} \left(
  e\varepsilon_{\mu\nu\rho}\,\xi^\nu {\cal P}^{\rho\,A}\,
 +{\cal V}^{\cN A}
 g_{\mu\nu}\nabla_{\cN}\xi^{\nu} \right)
 \;,
\nonumber \\
\Delta_{\xi} B_{\mu\,\cM}  &=& -e\varepsilon_{\mu\nu\rho}
\Big(
g^{\rho\lambda}\, { \cD}^\nu\left( g_{\lambda\sigma} \nabla_{\cM} \xi^\sigma \right) 
-  \xi^\nu  \widehat{J}^\rho{}_{\cM}
\Big)
 \;,
 \label{ext_diff}
\eea
where the variation of $B_{\mu\,\cM}$ is given in terms of  the current $\widehat{J}^\rho{}_{\cM}$ introduced in (\ref{BIR2})
and most compactly expressed via the general covariant variation $\Delta B_{\mu\,\cM}$ introduced in (\ref{DeltaB}).
With (\ref{DeltaB}), (\ref{chrisexpl}), and the explicit form of $\delta_\xi A_\mu{}^\cM$ it is
straightforward to verify that the variation $\delta_\xi B_{\mu\,\cM} $ is uniquely determined and
compatible with the constraints (\ref{sectionB}) this connection satifies.
The external diffeomorphisms (\ref{ext_diff}) take the expected form for the frame fields $e_\mu{}^a$, $\cM_{\cM\cN}$.
In contrast, for the gauge connections $A_{\mu}{}^{\cM}$, $B_{\mu\,\cM}$, they relate only on-shell to
the standard diffeomorphism transformation of gauge fields.

\subsection{Closure of the supersymmetry algebra}

Let us now move on to the fermionic fields and the supersymmetry algebra. 
In addition to the bosonic fields introduced in section \ref{exgeo}, the supersymmetric completion of the  ${\rm E}_{8(8)}$ exceptional field theory contains the following spinor fields: sixteen gravitinos ${\psi_{\mu}}^I$ as well as 128 matter fermions $\chi^{\dot{A}}$, transforming
in the vector and spinor representation of SO(16), respectively. 
With respect to generalized diffeomorphisms, they transform as scalar densities with half-integer weights given in Table~\ref{weights}.
We are working in the Majorana representation and mostly minus signature, i.e.\ spinors are taken to be real and SO(1,2) gamma matrices $\gamma_\mu$ purely imaginary, c.f.~\cite{Nicolai:2001sv} for our spinor conventions.
In particular, we use $\gamma_{\mu\nu\rho}=-ie\varepsilon_{\mu\nu\rho}$\,.

In this section, we present the supersymmetry transformation rules
\bea
\delta_\epsilon e_{\mu}{}^a&=&i\bar{\epsilon}^I\gamma^a\psi_{\mu}^I\;,\qquad \cV^{-1}\delta_\epsilon \cV~=~ \Gamma^I_{A\dot{A}}\bar{\chi}^{\dot{A}}\epsilon^I Y^A\;,
\nonumber\\
\delta_\epsilon \psi^I_{\mu}&=&\cD_{\mu}\epsilon^I+2\cV^\cM{}_{IJ}\nabla_\cM(i\gamma_{\mu}\epsilon^J)+2\cV^\cM{}_{IJ}\, i\gamma_{\mu}\nabla_\cM\epsilon^J\;,\nonumber\\
\delta_\epsilon\chi^{\dot{A}}&=&\frac{i}{2}\gamma^{\mu}\epsilon^I\Gamma^I_{A\dot{A}}\hat{\cP}^{A}_{\mu}-2\cV^{\cM}{}_{A}\Gamma^I_{A\dot{A}}\nabla_\cM\epsilon^I\;,\nonumber\\
\delta_{\epsilon}A_{\mu}{}^{\cM}&=&-4 \,\cV^{\cM}{}_{IJ}\,{\bar{\epsilon}}^I{\psi_{\mu}}^J+2 \Gamma{}^I_{A\dot{A}}\cV^{\cM}{}_A\,{\bar{\epsilon}}^I i\gamma_{\mu}\chi^{\dot{A}}\;,\nonumber\\
\Delta_\epsilon B_{\mu\,\cM}&=&-2 (\nabla_{\cM}\bar{\epsilon}^I \psi_{\mu}^I-\bar{\epsilon}^I \nabla_{\cM} \psi_{\mu}^I)+\,e\,\varepsilon_{\mu\nu\rho}g^{\rho\sigma}\nabla_\cM(\bar{\epsilon}^Ii\gamma^{\nu}\psi_{\sigma}^I)\;,
\label{susyvar}
\eea
and show its algebra closes into generalized diffeomorphisms and gauge transformations.
The bosonic transformationss (first and fourth line) precisely coincide with the supersymmetry transformations of standard D=3 supergravity \cite{Marcus:1983hb,Nicolai:2001sv} with all fields now living on the exceptional space-time. The fermionic transformation rules on the other hand have been modified w.r.t.\ the three-dimensional theory with the addition of term containing internal covariant derivatives $\nabla_{\cal M}$ introduced in section~\ref{subsec:spin}. 
As in higher dimensions, the supersymmetry transformation rules only carry specific projections of these covariant derivatives,
such that the undetermined part in the SO(16) connection ${\cal Q}_{\cM}{}^{IJ}$ drops out.
The supersymmetry variations of the gauge connection $B_{\mu\,\cM}$ finally have no analogue in the three-dimensional theory and are entirely determined from closure of the supersymmetry algebra. Although its r.h.s.\ is such that not all undetermined parts of the SO(16) connection ${\cal Q}_{\cM}{}^{IJ}$ drop out, these terms precisely cancel the corresponding contributions from the 
Christoffel connection in the covariant variation (\ref{DeltaB}) on the l.h.s.. The resulting variation $\delta_\epsilon B_{\mu\,\cM}$  is uniquely determined and compatible with the constraints (\ref{sectionB}) this field has to satisfy.

As a first test, we
use this ansatz to calculate the commutator of two supersymmetry transformations on the dreibein $e_\mu{}^a$ to obtain
\bea
\left[\delta_{\epsilon_1}, \delta_{\epsilon_2}\right]e_{\mu}{}^a&=&e\bar{\epsilon}_2^I\gamma^a\left(\cD_{\mu}\epsilon^I+2\cV^{\cM}{}_{IJ}\nabla_\cM(i\gamma_{\mu})\epsilon^J+4\cV^{\cM}{}_{IJ} i\gamma_{\mu}\nabla_\cM\epsilon^J\right)-(1\leftrightarrow2)
\nonumber\\
&=&\mathcal{D}_{\mu}\left(\bar{\epsilon}^I_2\,  i\gamma^a \, \epsilon^I_1\right)-4 \cV^{\cM}{}_{IJ}\bar{\epsilon}^I_2 \, \epsilon^J_1\nabla_{\cM}e_{\mu}{}^a
+\nabla_\cM\left( -4\cV^\cM{}_{IJ}\bar{\epsilon}^I_2 \, \epsilon^J_1\right)e_{\mu}{}^a
\nonumber\\
&&{}-4\,\cV^{\cM}{}_{IJ}\left(\bar{\epsilon}^I_2\,  \gamma^{ab}\, \nabla_\cM \epsilon^J_1-\nabla_\cM\bar{\epsilon}^I_2\,  \gamma^{ab} \, \epsilon^J_1\right)e_{\mu\,b}
\nonumber\\
&\equiv&
\mathcal{D}_{\mu}(\xi^{\nu}e_{\nu}{}^a) +\Lambda^\cM\partial_\cM e_{\mu}{}^a+\partial_\cM \Lambda^\cM e_{\mu}{}^a+\tilde{\Omega}^{ab}e_{\mu\,b}\;.
\label{dreiclose}
\eea
The first term reproduces the action of covariantized external diffeomorphisms, the second and third term describe the action of internal generalized diffeomorphisms on the dreibein, and the last term is an SO(1,2) Lorentz transformation, with the respective parameters given by
\bea
\xi^{\mu}&=&i\bar{\epsilon}^I_2\,  \gamma^{\mu} \, \epsilon^I_1\;,\nonumber\\
\Lambda{}^\cM&=&-4 \,\cV^{\cM}{}_{IJ}\bar{\epsilon}^I_2\epsilon^J_1\;,\nonumber\\
\tilde{\Omega}^{ab}&=&-4\,\cV^{\cM}{}_{IJ}\left(\bar{\epsilon}^I_2\,  \gamma^{ab}\, \nabla_\cM \epsilon^J_1-\nabla_\cM\bar{\epsilon}^I_2\,  \gamma^{ab} \, \epsilon^J_1\right) + \Lambda^\cM\,\omega_{\cM}{}^{ab}
\;.
\label{LambdaSusy}
\eea

Similarly, one can show closure of the supersymmetry algebra on the 248-bein.
Using $\eqref{susyvar}$, we find the commutator
 \begin{align}
\mathcal{V}^{\cM}{}_B\left[\delta_{\epsilon_1}, \delta_{\epsilon_2}\right]\mathcal{V}_{\cM}{}^{KL}&=
\left(-\frac{i}{2}\cP_{\mu}{}^C\Gamma^J_{C\dot{A}}\bar{\epsilon}_1^J\gamma^\mu-2\cV^{\cN}{}_C\Gamma^J_{C\dot{A}}\nabla_{\cN}\bar{\epsilon}_1^J\right)\epsilon_2^I\Gamma^{I}_{A\dot{A}}(Y^A)^{KL}{}_B\nonumber\\
&\quad-(1\leftrightarrow2)\nonumber\\
&=\xi^{\mu}\cP_{\mu}{}^A\,(Y^A)^{KL}{}_B+60\,\mathcal{V}^\cM{}_B\P^{\cN}{}_\cM{}^{\cK}{}_{\cL}\mathcal{V}_{\cN}{}^{KL}\nabla_{\cK}\Lambda^{\cL}\nonumber\\
&\quad -2\,\mathcal{V}^{\cM}{}_B\left(\nabla_\cN \bar{\epsilon}^I_2\epsilon^I_1-\bar{\epsilon}^I_2 \nabla_\cN \epsilon^I_1\right) f^{\cN}{}^{\cP}{}_{\cM}\mathcal{V}_{\cP}{}^{KL}
\;,
\label{closV1}
\end{align}
with the adjoint projector from (\ref{Padj}).
We recognize the first term as the action of external diffeomorphisms on the 248-bein. 
The second term reproduces the action (\ref{GLie}) of a generalized internal diffeomorphism
with parameter $\Lambda^\cL$ when parametrized covariantly as in (\ref{SigSig}) (note that
the transport term $\Lambda^\cN \nabla_{\cN} \mathcal{V}_{\cM}{}^{KL}$ vanishes due to the
vielbein postulate (\ref{GVP})).
The last term thus describes the covariantized E$_{8(8)}$ rotation from which we read off the parameter
$\tilde{\Sigma}_\cN$
\bea
\tilde{\Sigma}_\cN&=&
-2\left(\nabla_\cN \bar{\epsilon}_2^I\epsilon_1^I-\bar{\epsilon}_2^I \nabla_\cN \epsilon_1^I\right)
\;.
\label{StildeSusy}
\eea
As a consistency check, it is straightforward to verify that although the expression for the
parameter (\ref{StildeSusy}) carries the full internal SO(16) spin connection ${\cal Q}_{\cN}{}^{IJ}$
(including its undetermined parts), its form is such that the constrained parameter 
$\Sigma_{\cN}=\tilde{\Sigma}_\cN+\Gamma_{\cN,\cM}\Lambda^{\cM}$ 
which actually appears in the rotation term of (\ref{GLie}) is uniquely determined 
(with the undetermined part from ${\cal Q}_{\cN}{}^{IJ}$ cancelling the undetermined part from $\Gamma_{\cN,\cM}$)
and moreover satisfies the required constraints (\ref{sectionB}).

Also on the gauge field $A_{\mu}{}^{\cM}$ we obtain closure of the supersymmetry algebra by a standard calculation
which gives the explicit result
\bea
\left[\delta_{\epsilon_1}, \delta_{\epsilon_2}\right]A_{\mu}{}^{\cM}&=&-4\,\cV^{\cM}{}_{IK}\,\bar{\epsilon}^I_2 \left(
\cD_{\mu}\epsilon_1^K+2\,\cV^{\cN}{}_{KJ}\nabla_{\cN}(i\gamma_{\mu})\epsilon_1^J+4\cV^{\cN}{}_{KJ}\,i\gamma_{\mu}\nabla_{\cN}\epsilon_1^J \right)\nonumber\\
&&+2\,\Gamma^{I}_{A\dot{A}}\cV^{\cM}{}_A\bar{\epsilon}_2^Ii\gamma_{\mu}\left(\frac{i}{2}\,\gamma_{\nu}\epsilon_1^J\Gamma^J_{B\dot{A}}\cP^{\nu\,B}-2\,\cV^{\cN}{}_B\Gamma^{J}_{B\dot{A}}\nabla_{\cN}\epsilon_1^J\right)-(1\leftrightarrow 2)
\nonumber\\
&=& \cD_{\mu}\Lambda^{\cM}+\nabla_{\cN}\left(-16i\,\cV^{\cM}{}_{K(I}\cV^{\cM}{}_{J)K}\bar{\epsilon}_2^I\gamma_{\mu}\epsilon_1^J\right)
\nonumber\\
&&{}
+8i\,f^{\cM \cN}{}_{\cK}\cV^{\cK}{}_{IJ}\left(\bar{\epsilon}_2^I\gamma_{\mu}\nabla_{\cN}\epsilon_1^J-\nabla_{\cN}\bar{\epsilon}_2^I\gamma_{\mu}\epsilon_1^J\right)\nonumber\\
&&-2i\,e\varepsilon_{\mu\nu\rho}\cV^{\cM}{}_A\cP^{\rho \, A}\bar{\epsilon}_2^Ii\gamma^{\nu}\epsilon_1^I+4\cV^{\cM}{}_A\cV^{\cN}{}_A\xi^a\nabla_{\cN}e_{\mu}{}^a-4\cV^{\cM}{}_A\cV^{\cN}{}_A \nabla_{\cN}\xi_{\mu}
\nonumber\\
&=&\cD_{\mu}\Lambda^{\cM} -2{\cal V}^{\cM A} \left(
  e\varepsilon_{\mu\nu\rho}\,\xi^\nu {\cal P}^{\rho\,A}\,
 +{\cal V}^{\cN A}
 g_{\mu\nu}\nabla_{\cN}\xi^{\nu} \right)\nonumber \\
&&+\partial_{\cN}{\Xi_{\mu\,3875}}{}^{(\cM\cN)}+f^{\cM\cN}{}_{\cK}\,\Xi_{\mu\,\cN}{}^{\cK}+\eta^{\cM \cN}\Xi_{\mu\,\cN}
\label{closeA}
\eea
with the parameters $\Lambda^\cM$ and $\xi^\mu$
from (\ref{LambdaSusy}) and the shift parameters $\Xi_{\mu}$ of the last line defined as
\bea
\Xi_{\mu\,\cN}&=& -2 \partial_{\cN}\xi_{\mu}\;,
\nonumber\\
{\Xi_{\mu\,3875}}^{(\cM\cN)}&=&
-16\,\cV^{\cM}{}_{IK}\cV{}^{\cN}{}_{KJ}\bar{\epsilon}_2^{(I}i\gamma_{\mu}\epsilon_1^{J)}
-\cV^{\cM}{}_{IJ}\cV^{\cN}{}_{IJ}\bar{\epsilon}_2^{K}i\gamma_{\mu}\epsilon_1^{K}
\;,\nonumber\\
{\Xi_{\mu}}_{\cN}{}^{\cK}
&=&-8\,\cV^{\cK}{}_{I J}\left(\nabla_\cN\bar{\epsilon}_2^{I}i\gamma_{\mu}\epsilon_1^{J}
-\bar{\epsilon}_2^{I}i\gamma_{\mu}\nabla_\cN\epsilon_1^{J}\right)\nonumber\\
&&+\Gamma_{\cN, \cM}\left(\,{\Xi_{\mu\,3875}}^{(\cM\cK)}-2\eta^{\cM \cK}\xi_{\mu}\right)\;
\;,
\label{shiftsSusy}
\eea
corresponding to the shift symmetries (\ref{shiftAB}) discussed above. The fact that ${\Xi_{\mu\,3875}}^{(\cM\cN)}$ lives in
${\bf 3875}$ representations is an immediate consequence of its specific form 
\bea
{\Xi_{\mu\,3875}}^{(\cM\cN)}=-16\,\cV^{\cM}{}_{IK}\cV{}^{\cN}{}_{KJ}\,\xi_{\mu\,IJ}
\;,\qquad
\xi_{\mu\,IJ}\equiv
i\bar{\epsilon}_2^{(I}\gamma_{\mu}\epsilon_1^{J)}-\frac1{16}\,\delta_{IJ}\,\xi_\mu
\;,
\eea
with a parameter $\xi_{\mu\,IJ}$ in the ${\bf 135}$ of SO(16), combined with the fact that the tensor product
of two adjoint representations (\ref{adjadj}) contains only a single representation ${\bf 135}$ of SO(16) which lives
within the ${\bf 3875}$ representation of E$_{8(8)}$\,.
Moreover, the last term in (\ref{shiftsSusy}) carrying the Christoffel connection
ensures that the parameter ${\Xi_{\mu}}_{\cN}{}^{\cK}$ does not carry any of the undetermined parts
of the SO(16) connection ${\cal Q}_\cN{}^{IJ}$ and furthermore is constrained in its first index,
as required by the shift symmetries (\ref{shiftAB}).

We have at this point fully determined the supersymmetry algebra 
\bea
\left[\delta_{\epsilon_1}, \delta_{\epsilon_2}\right] &=&
\delta_\xi + \delta_{\tilde{\Omega}}+\delta_\Lambda + \delta_\Sigma
+ \delta_\Xi
\;,
\label{susyalgebra}
\eea
with parameters given in (\ref{LambdaSusy}), (\ref{StildeSusy}), (\ref{shiftsSusy}).
As a consistency check of the construction it remains to verify that the algebra
closes in the same form on the constrained connection $B_{\mu\,\cM}$.
This computation is greatly facilitated by the notation of the
covariant variation (\ref{DeltaB}) in terms of which its supersymmetry variation takes the
covariant form (\ref{susyvar}). To lowest order in fermions, the supersymmetry algebra on $B_{\mu\,\cM}$
is given by
\bea
 \left[\delta_{\epsilon_1}, \delta_{\epsilon_2}\right] B_{\mu\,\cM}
 &=&
 2\,\delta\undersym{{}_{\epsilon_1}\,\Delta_{\epsilon_2}}\,B_{\mu\, \cM}
 +2\,\Gamma_{\cM,\cN}\,\delta\undersym{{}_{\epsilon_1}\,\delta_{\epsilon_2}}\,A_{\mu}{}^{\cN}
 \;.
\eea
For the second term we may use the closure of the algebra on the vector fields $A_\mu{}^\cM$
established above. 
The first term after some calculation yields 
\bea
2\,\delta\undersym{{}_{\epsilon_1}\,\Delta_{\epsilon_2}}\,B_{\mu\, \cM}&=& \Delta_{\Lambda,\Sigma}B_{\mu\, \cM}+\Delta_{\xi}B_{\mu\, \cM}\nonumber\\
&&+2\nabla_{(\cM}\tilde{\Xi}_{\mu\,\cN)}{}^{\cN}+\varepsilon_{\mu\nu\rho}\cR_{\cM \cN}{}^{\nu\rho}\Lambda^\cN\nonumber\\
&&+8\,\cV^{\cN}{}_{IJ}\left([\nabla_{\cM},\nabla_{\cN}]\bar{\epsilon}_2^Ii\gamma_{\mu}\epsilon_1^J-\bar{\epsilon}_2^Ii\gamma_{\mu}[\nabla_{\cM},\nabla_{\cN}]\epsilon_1^J\right)
\;,
\label{SusycloseB}
\eea
with the parameters given in (\ref{LambdaSusy}), (\ref{StildeSusy}), (\ref{shiftsSusy}) and the covariant combination
\bea
 \tilde{\Xi}_{\mu\,\cN}{}^{\cK}&=&-8\,\cV^{\cK}{}_{IJ}\left(\nabla_\cN\bar{\epsilon}_2^{I}i\gamma_{\mu}\epsilon_1^{J}
-\bar{\epsilon}_2^{I}i\gamma_{\mu}\nabla_\cN\epsilon_1^{J}\right)\;,\nonumber\\
&=&{\Xi_{\mu}}_{\cN}{}^{\cK}-\Gamma_{\cN, \cM}\left(\,{\Xi_{\mu\,3875}}^{(\cM\cK)}-2\eta^{\cM \cK}\xi_{\mu}\right)\;.
\eea
The first line of $\eqref{SusycloseB}$ reproduce the covariant variation of $B_{\mu\, \cM}$ under generalized internal and external diffeomorphisms. For the supersymmetry algebra to close, the second and third line of $\eqref{SusycloseB}$ must reproduce the shift symmetries
\bea
\Delta_{\Xi}B_{\mu\, \cM}&=&\delta_{\Xi}B_{\mu\, \cM}-\Gamma_{\cM, \cN}\,\delta_{\Xi}A_{\mu}{}^{\cM}\;,\nonumber\\
&=&2\,\nabla_{(\cM}\tilde{\Xi}_{\mu\, \cN)}{}^{\cN}+2\Gamma_{[\cN \cM]}{}^{\cP}\tilde{\Xi}_{\mu \, \cP}{}^\cN-\Gamma_{\cP [\cN}{}^\cP\tilde{\Xi}_{\mu\, \cM]}{}^\cN\nonumber\\
&&\left(\partial_{\cN}\Gamma_{\cN, \cP}-\Gamma_{\cN \cP}{}^\cQ\Gamma_{\cM, \cQ}\right)\left({\Xi_{\mu\,3875}}^{\cN\cP}-2\eta^{\cN \cP}\xi_{\mu}\right)\;,\nonumber\\
&=& 2\,\nabla_{(\cM}\tilde{\Xi}_{\mu\,\cN)}{}^{\cN}+\varepsilon_{\mu\nu\rho}\cR_{\cM \cN}{}^{\nu\rho}\Lambda^\cN\nonumber\\
&&+8\,\cV^{\cN}{}_{IJ}\left([\nabla_{\cM},\nabla_{\cN}]\bar{\epsilon}_2^Ii\gamma_{\mu}\epsilon_1^J-\bar{\epsilon}_2^Ii\gamma_{\mu}[\nabla_{\cM},\nabla_{\cN}]\epsilon_1^J\right)\;,
\eea
where we have obtained the last equality with the use of the following identity
\bea
\left(2\partial_{[\cM}\Gamma_{\cN], \cP}-\Gamma_{\cM,\cL}\Gamma_{\cN, \cQ}f^{\cL \cQ}{}_{\cP}\right)\left(\cV^{(\cN}{}_{IK}\cV^{\cP)}{}_{KJ}+\frac{1}{8}\cV^{\cN}{}_A\cV^{\cP}{}_A\,\delta_{I J}\right)&&
\nonumber\\
-\cV^{\cN}{}_{I K} \left(2\partial_{[\cM}Q_{\cN]}{}^{KJ}+2Q_{[\cM}{}^{KL}Q_{\cN]}{}^{LJ}\right)&=&0\;.
\eea
This is reminiscent of standard Riemannian geometry, where the curvature of the Christoffel symbols is the curvature of the spin connection
\bea
R_{\mu\nu}{}^{\rho\sigma}\,[\Gamma]=R_{\mu\nu}{}^{ab}\,[\omega]\,e_{a}{}^\rho e_{b}{}^\sigma\;,
\eea
albeit here, in a projected fashion.

This proves the closure of the supersymmetry algebra on $B_{\mu \, \cM}$
\bea
\left[\delta_{\epsilon_1}, \delta_{\epsilon_2}\right] B_{\mu\,\cM}
 &=& \delta_{(\Lambda,\Sigma)}B_{\mu\,\cM}+\delta_{\xi}B_{\mu\,\cM}+\delta_{\Xi}B_{\mu\,\cM}\;,
\eea
and concludes the discussion on the consistency of the supersymmetry algebra $\eqref{susyalgebra}
$.

%%%%%%%%%%%%%%%%%%%%%%%%%%%%%%%%%%%%%%%%%%%%%%%%%%%%%
%%%%%%%%%%%%%%%%%%%%%%%%%%%%%%%%%%%%%%%%%%%%%%%%%%%%%
\section{Action}
\label{sec:action}
%%%%%%%%%%%%%%%%%%%%%%%%%%%%%%%%%%%%%%%%%%%%%%%%%%%%%
%%%%%%%%%%%%%%%%%%%%%%%%%%%%%%%%%%%%%%%%%%%%%%%%%%%%%

Having introduced fermion fields and supersymmetry transformation laws, we now have at our disposal
all the necessary tools to construct the fermionic completion of the 
${\rm E}_{8(8)}$ bosonic Lagrangian constructed in \cite{Hohm:2014fxa}. 
We start by giving a brief review of the bosonic Lagrangian in the form most suited for
the coupling of fermions before presenting its supersymmetric completion.

%%%%%%%%%%%%%%%%%%%%%%%%%%%%%%%%%%%%%%%%%%%%%%%%%%%%%
\subsection{The bosonic theory}
\label{subsec:bosonic}
%%%%%%%%%%%%%%%%%%%%%%%%%%%%%%%%%%%%%%%%%%%%%%%%%%%%%

Let us start by a brief review of the action of bosonic E$_{8(8)}$ exceptional field theory
following \cite{Hohm:2014fxa}\footnote{As mentioned above, in this paper we use the metric signature $(+--)$. 
Some signs in the present Lagrangian thus differ from the ones in \cite{Hohm:2014fxa}
which was given in mostly plus signature.
 }
however adapted to the further coupling of fermions,
in particular using the internal frame field (the 248-bein) from (\ref{e8so16}).
The bosonic field content has been given and discussed in (\ref{fields_bos}).
The action of bosonic ${\rm E}_{8(8)}$ exceptional field theory is given by\footnote{
As usual, the integral over the 248 internal coordinates is to be taken in a formal sense
since the section constraint (\ref{section}) remains to be imposed by hand and
eliminates the field dependence on most of these coordinates.
}
\bea
{S}_{\rm bos} &=&
\int d^3x\, d^{248}Y\;
\left(
{\cal L}_{\rm EH}+ {\cal L}_{\rm scalar}+{\cal L}_{\rm top}+{\cal L}_{\rm pot}\right)
\;,
\label{action_bos}
\eea
where each term is separately invariant under generalized internal diffeomorphisms (\ref{GLie}).
The Einstein-Hilbert Lagrangian is given by the Ricci scalar (\ref{REH})
obtained from contraction of the improved Riemann tensor 
\bea
{\cal L}_{\rm EH} &=& -e\,e_a{}^\mu e_b{}^\nu\, \widehat{\cal R}_{\mu\nu}{}^{ab}
\;,
\eea
where $e$ denotes the determinant of the dreibein $e_\mu{}^a$.
The scalar kinetic term in (\ref{action_bos}) is given by
\bea
{\cal L}_{\rm scalar}&=& -\frac1{240} \,{D}_\mu {\cal M}_{\cM\cN}{D}^\mu {\cal M}^{\cM\cN}
~=~  {\cal P}_\mu{}^A \,{\cal P}^{\mu\,A}
\;,
\eea
where we have used the expression
\bea
{\cal M}^{\cK\cP}{D}_\mu {\cal M}_{\cP\cL} &=& 
2\,f^{{\cal M}{\cal K}}{}_{\cal L}\,{\cal V}_{\cal M}{}^A {\cal P}_\mu{}^A
\;,
\label{JinP}
\eea
of the scalar currents (with covariant derivatives from (\ref{covD})) 
in terms of the E$_{8(8)}$ structure constants $f^{{\cal M}{\cal K}}{}_{\cal L}$ and
the coset currents (\ref{Qmu}). 
The topological term in (\ref{action_bos}) carries the non-abelian Chern-Simons couplings
of the gauge connections according to
\bea
{\cal L}_{\rm CS} &=&
-\frac12\,\varepsilon^{\mu\nu\rho}\,\Big( 
{\cal F}_{\mu\nu}{}^\cM B_{\rho}{\,}_\cM
-f_{\cK\cL}{}^\cN \partial_\mu A_\nu{}^\cK \partial_\cN A_\rho{}^\cL
-\frac23\,f^\cN{}_{\cK\cL} \partial_\cM\partial_\cN A_{\mu}{}^\cK A_\nu{}^\cM A_{\rho}{}^\cL
\nonumber\\
&&{}
\qquad\qquad
-\frac13 \,f_{\cM\cK\cL} f^{\cK\cP}{}_\cQ f^{\cL\cR}{}_\cS\,A_\mu{}^\cM \partial_\cP A_\nu{}^\cQ \partial_\cR A_\rho{}^\cS
\Big) 
\;.
\label{CS}
\eea
Its covariance becomes manifest upon spelling out its variation as
\bea
\delta\cL_{\rm CS}&=&
-\frac12\,\varepsilon^{\mu\nu\rho}\Big({\cal F}_{\mu\nu}{}^{\cM}\,\Delta B_{\rho \,\cM}
+\left(
\widetilde{\cal G}_{\mu\nu\,\cM}
+2\,f_{\cM\cN}{}^\cK\,\nabla_\cK {\cal F}_{\mu\nu}{}^{\cN}
\right)
\delta A_\rho{}^{\cM} \Big)
\;,
\eea
with the covariant field strengths from (\ref{FG}), (\ref{deltaFG}) 
and the general covariant variation introduced in (\ref{DeltaB}). 
As anticipated above, we note that the two-form contributions to the field
strengths ${\cal F}$ and ${\cal G}$ (whose explicit form has been suppressed in (\ref{FG}))
drop out from this expression due to the section constraint.
Moreover, the contributions to the Christoffel connection in $\nabla_{\cK}$ that are
left undetermined by the vanishing torsion condition cancel in this expression
against the corresponding contributions in $\Delta B_{\rho \,\cM}$.

Finally, the last term in (\ref{action_bos}) carries only derivatives in the internal coordinates
and is explicitly given by
\bea
{\cal L}_{\rm pot} &=& -e\,V\;,
\eea
with the `potential' $V$ given by
\bea
  V & \equiv &
 \frac{1}{4}{j}_\cM{}^\cR{j}_\cN{}^\cS
\left({\cal M}^{\cM\cN} 
\eta_{\cR\cS}
-2{\cal M}^{\cK\cL}  
 f_{\cR\cL}{}^\cN f_{\cS\cK}{}^\cM+2\delta_\cR{}^{\cN}\delta_{\cS}{}^{\cM}
\right)
  \label{VinJ}\\  &&{}
  -\frac{1}{2}g^{-1}\partial_\cM g\,{\cal M}^{\cM\cN}f_{\cN\cK}{}^{\cP} {j}_{\cP}{}^{\cK}
  -\frac{1}{4}  {\cal M}^{\cM\cN}g^{-1}\partial_\cM g\,g^{-1}\partial_\cN g
  -\frac{1}{4}{\cal M}^{\cM\cN}\partial_\cM g^{\mu\nu}\partial_\cN g_{\mu\nu}\;,
\nonumber
\eea
in terms of the internal current ${j}_\cM{}^\cN$ defined as
\bea
{\cal M}^{\cK\cP}\partial_\cM {\cal M}_{\cP\cL} &\equiv& {j}_\cM{}^\cN\,f_{\cN\cL}{}^\cK
\nonumber\\
&=&
2\,f_{\cN\cL}{}^\cK
p_{\cM}{}^A\cV^{\cN}{}_{A}~=~
 -f_{\cN\cL}{}^\cK \,\tilde\Gamma_{\cM,\cP}\left({\cal M}^{\cP\cN}+\eta^{\cP\cN}\right)
\;,
\label{currJ2}
\eea
where we have used $\eqref{chrisexpl}$ in the last equality. The scalar potential (\ref{VinJ}) then takes the manifestly covariant form
\bea
V  &=& \cR  -\frac{1}{4}{\cal M}^{\cM\cN}\nabla_\cM g^{\mu\nu}\nabla_\cN g_{\mu\nu} +\nabla_{\cM}I^{\cM}
\;,
\label{scapot}
\eea
with the internal curvature scalar $\cR$ from $\eqref{curvsca}$ 
and up to a boundary contribution $I^{\cM}$ of weight $\lambda_I=-1$\,.
This completes the definition of the bosonic Lagrangian.

Let us finally present the vector field equations in a manifestly covariant form.
Parametrizing the variation of the Lagrangian (\ref{action_bos}) w.r.t.\ the gauge fields as
\bea
\delta\cL&=&\varepsilon^{\mu\nu\rho}\left({\cal E}_{\mu\nu}^{(A)\,\cM}\,\Delta B_{\rho \,\cM}
+\widehat{\cal E}_{\mu\nu\,\cM}^{(B)}\,\delta A_\rho{}^{\cM} \right)
\;,
\label{variation}
\eea
with the general covariant variation of (\ref{DeltaB}),
the Chern-Simons couplings together with the minimal couplings in the covariant derivatives give rise to
duality equations relating the field strengths to matter currents according to
\bea
{\cal E}_{\mu\nu}^{(A)\,\cM} &\equiv& 
-\frac12\,{\cal F}_{\mu\nu}{}^\cM
+\frac{1}2\, e\,\varepsilon_{\mu\nu\rho}\,  
{j}^\rho{}^\cM
\;,
\nonumber\\
\widehat{\cal E}_{\mu\nu\,\cM}^{(B)} 
&=&
-\frac12\,\widetilde{\cG}_{\mu\nu\,{\cal M}} 
+\frac12\,f_{\cM \cN}{}^\cK\nabla_\cK{\cal F}_{\mu\nu}{}^{\cN}
-\frac{1}{2}\, e\,\varepsilon_{\mu\nu\rho}\left(
f_{\cM \cN}{}^\cK\nabla_\cK j^{\rho\,\cN}
+\widehat{J}^\rho{}_{\cM} \right)
\;,
\eea
with the covariant field strengths from (\ref{FG}), (\ref{deltaFG}) 
and the current $\widehat{J}^\rho{}_{\cM}$ from (\ref{BIR2}).
Let us stress that the equations of motion do not imply
the full vanishing ${\cal E}_{\mu\nu}^{(A)\,\cM}=0$ due to the fact
that the corresponding variation (\ref{variation}) is a variation w.r.t.\ a constrained gauge connection
subject to the section constraint (\ref{sectionB}).
In other words, the equations of motion only imply the weaker projected equation
\bea
{\cal E}_{\mu\nu}^{(A)\,\cM} &=& {\cal O}_{\mu\nu}{}^{\cM}
\;,
\eea
where ${\cal O}_{\mu\nu}{}^{\cM}$ vanishes when contracted with a field satisfying the section constraints (\ref{sectionB}).

\subsection{Supersymmetric Lagrangian}

We can now present the supersymmetric completion of the bosonic action (\ref{action_bos}).
The fermionic field content comprises the gravitinos $\psi_\mu{}^I$ and spin 1/2 fermions $\chi^{\dot{A}}$ 
transforming in the fundamental vector ${\bf 16}$ and spinor ${\bf 128_c}$
representations of SO(16), respectively.
The full $\rm{E}_{8(8)}$ Lagrangian is given by
\bea
e^{-1}\cL&=&-\widehat{\R}+g^{\mu\nu}\cP_{\mu}{}^{A}\cP_{\nu}{}^A+e^{-1}\,\cL_{\rm top}-V\nonumber\\
&&+2i\, \gamma^{\mu\nu\rho}\bar{\psi}_{\lambda}^I\cD_{\mu}\psi_{\nu}^I
-2i\,\bar{\chi}^{\dot{A}}\gamma^{\mu}\cD_{\mu}\chi^{\dot{A}}
-2\,\bar{\chi}{}^{\dot{A}}\gamma^{\mu}\gamma^{\nu}\psi_{\mu}^I\Gamma^I_{A\dot{A}}\cP_{\nu}{}^A\nonumber\\
&&
+e^{-1}\,{\cal L}_{\rm quartic}
+8\,\cV{}^\cM{}_{IJ}\bar\psi_{\mu}^I\gamma^{\mu\nu}\nabla_{\cM}\psi_{\nu}^J
-8i \,\cV{}^\cM{}_{A}\, \Gamma^I_{A\dot{A}}\bar\psi_{\mu}^I\nabla_{\cM}(\gamma^{\mu}\chi^{\dot{A}})
\nonumber\\
&&
-2\,\cV{}^\cM{}_{IJ}\, \Gamma^{IJ}_{\dot{A}\dot{B}}\bar{\chi}^{\dot{A}}\nabla_{\cM}\chi^{\dot{B}}
\;.
\label{completelag}
\eea
The first line is the bosonic Lagrangian $\eqref{action_bos}$. The terms in the second line are obtained via a direct uplift (and proper covariantization) from $D=3$ maximal supergravity \cite{Marcus:1983hb,Nicolai:2001sv}: a Rarita-Schwinger term for the gravitinos $\psi_{\mu}^I$, a kinetic term for the 128 matter fermions $\chi^{\dot{A}}$, and the Noether coupling between the coset current $\cP_{\mu}{}^A$ and the fermions. The three last terms of (\ref{completelag}) 
carrying internal covariant derivative $\nabla_{\cM}$ 
have been added to ensure invariance of the Lagrangian under supersymmetry transformations. 
After proper Scherk-Schwarz reduction of the Lagrangian~\cite{Hohm:2014qga}, 
these terms provide the Yukawa couplings of the
gauged three-dimensional supergravity.
Finally, ${\cal L}_{\rm quartic}$ denotes the quartic fermion terms. 
We expect these to coincide with the corresponding terms
of the three-dimensional theory~\cite{Marcus:1983hb,Nicolai:2001sv}
\bea
e^{-1}\,{\cal L}_{\rm quartic} &=&
-\frac12 \,\Big( \overline{\chi} \gamma_\rho \Gamma^{IJ} 
\chi \left( \overline{\psi}{}_\mu^I
 \gamma^{\mu\nu\rho} \psi_\nu^J 
- \overline{\psi}{}_\mu^I \gamma^\rho \psi^{\mu J}\right)
+ \overline{\chi} \chi \, \overline{\psi}{}_\mu^I \gamma^\nu \gamma^\mu \psi_\nu^I \Big)
\nonumber\\
&&{} + \frac12\,\Big(  (\overline{\chi} \chi) (\overline{\chi} \chi) - 
 \frac1{12}\, \overline{\chi} 
\gamma^\mu \Gamma^{IJ}\chi\,\overline{\chi} \gamma_\mu \Gamma^{IJ}\chi \Big)
\;,
\eea
but as far as this paper is concerned we will only deal with fermions at quadratic order.

For the proof of invariance of (\ref{completelag}) under supersymmetry (\ref{susyvar}), 
we first note that all terms that do not carry internal derivatives cancel precisely as in the three-dimensional theory. 
Terms carrying internal derivatives arise in the bosonic sector from variation of the potential $V$ 
and the topological term ${\cal L}_{\rm top}$. In the fermionic sector, such terms arise from the  
corresponding terms in the supersymmetry transformations (\ref{susyvar}), from variation of the last three terms in (\ref{completelag}),
as well as from the modified integrability relations (\ref{comderextcov}), (\ref{intP1}).

We organise these terms according to their structure
\bea
\bar{\psi}\,\cD_{\mu}\nabla_{\cM}\epsilon\;, \quad 
\bar{\chi}\, \cD_{\mu}\nabla_{\cM}\epsilon\;, \quad 
\bar\psi\, \nabla_{\cM} \nabla_{\cN} \epsilon\;, \quad 
\bar\chi\, \nabla_{\cM} \nabla_{\cN} \epsilon
\label{sectvarlag}
\eea
and show that they cancel against the contributions from the bosonic Lagrangian. In the rest of this section, we will only focus on the last two types of terms in $\eqref{sectvarlag}$, which carry two internal derivatives and thus exhibit an interesting geometric structure of the internal space. The cancellation of the remaining terms is described in detail in appendix~\ref{app:susy}.

Let us start by collecting the terms in $\bar\psi \,\nabla_{\cM} \nabla_{\cN} \epsilon$
in the variation of the fermionic Lagrangian
\begin{align}
e^{-1}\delta\cL_{\rm ferm}  \Big|_{\bar\psi \nabla \nabla \epsilon} ~\longrightarrow~&\, 8i\Big(8\cV^{\cM}{}_{
IK}\cV{}^{\cN}_{KJ}+\cV^{\cM}{}_{A}\cV^\cN{}_{A} \delta_{I J}\Big)\,\bar{\psi}_{\mu}^I\gamma^{\mu}\left\{\nabla_{\cM},\nabla_{\cN}\right\}\epsilon^J\nonumber\\
&+8i\Big(8 \cV^{\cM}{}_{IK}\cV^{\cN}{}_{KJ}+\Gamma^{IJ}_{AB}\cV^{\cM}{}_{A}\cV^{\cN}{}_{B}\Big)\,\bar{\psi}_{\mu}^I\gamma^{\mu}\left[\nabla_{\cM},\nabla_{\cN}\right]\epsilon^J\;\nonumber\\
&+32i\,\bar\psi_{\mu}^I\,\cV^{\cM}{}_{IK}\cV^{\cN}{}_{KJ}\Big(\gamma^{\mu\nu}\nabla_{\cN}\gamma_{\nu}\nabla_{\cM}\epsilon^J+2\gamma^{\mu\nu}\nabla_{\cM}\gamma_{\nu}\nabla_{\cN}\epsilon^J
\nonumber\\
&\qquad\qquad\qquad\qquad\qquad
+\nabla_{\cM}(\gamma^{\mu\nu})\gamma_{\nu}\nabla_{\cN}\epsilon^J\Big)\nonumber\\
&+16i\,\cV^{\cM}{}_A(\Gamma^{I}\Gamma^{J})_{AB}\cV^{\cN}{}_B\, \bar\psi_{\mu}^I\,\nabla_{\cM}\gamma^{\mu}\nabla_{\cN}\epsilon^J\nonumber\\
&+32i\,\cV^{\cM}{}_{IK}\cV^{\cN}{}_{KJ}\bar\psi_{\mu}^I\Big(\gamma^{\mu\nu}\nabla_{\cM}\nabla_{\cN}\gamma_{\nu}+\frac12\nabla_{\cM}\gamma^{\mu\nu}\nabla_{\cN}\gamma_{\nu}\Big)\epsilon^J\;.
\end{align}
Upon use of the section constraints (\ref{sectionso16}) 
and together with the identity $\eqref{comintder}$, one can show that all the quadratic and linear terms in derivatives of $\epsilon$ vanish.
Then, the remaining terms cancel the first two lines of the variation of the scalar potential $\eqref{scapot}$ under a supersymmetry transformation (up to total derivatives)
\bea\label{varpot}
\delta_{\epsilon}(e V)&=& \frac12 \,e\, \Big(g^{\mu\nu}\cR-\frac14 g^{\mu\nu}\cM^{\cM \cN}\nabla_{\cM}g^{\rho\sigma}\nabla_{\cN}g_{\rho\sigma}+\nabla_{\cM}(\cM^{\cM \cN}\nabla_{\cN}g^{\mu \nu})\nonumber\\
&&\qquad\qquad+g^{\mu\rho}\nabla_{\cM}g_{\rho\sigma}\nabla_{\cN}g^{\sigma \nu}\cM^{\cM \cN}\Big)\,
\delta_{\epsilon} g_{\mu\nu}\nonumber\\
&&{}+e\,\Gamma^{I}{}_{A\dot{A}}\,\overline{\chi}^{\dot{A}} \epsilon^I \Big(\cR_{A}+\frac14 \Gamma^{I J}_{A B}\cV^{(\cM}{}_{B}\cV^{\cN)}{}_{I J}\nabla_{\cM}g^{\mu\nu}\nabla_{\cN}g_{\mu\nu}\Big)\;,
\eea
where for the cancellation we have used the following identity 
\bea
\gamma^{\mu\nu}\nabla_{\cM}\nabla_{\cN}\gamma_{\nu}+\frac12\,\nabla_{\cM}\gamma^{\mu\nu}\nabla_{\cN}\gamma_{\nu}&=&\frac12\,\nabla_{\cM}\nabla_{\cN}\gamma^{\mu}-\frac12\,g^{\mu\nu}\nabla_{\cM}\nabla_{\cN}\gamma_{\nu}\nonumber\\
&&-\frac14\,\cR_{\cM \cN}{}^{a b}\gamma^{\mu}\gamma_{a b}-\frac18\,\gamma^{\mu}\nabla_{\cM}g^{\nu \rho}\nabla_{\cN}g_{\nu \rho}\;.
\eea
The last line in (\ref{varpot}) then cancels against the corresponding terms from the 
variation of the fermionic Lagrangian
\begin{align}\label{chinabnabeps}
e^{-1}\delta\cL_{\rm ferm} \Big|_{\bar{\chi} \nabla \nabla \epsilon} ~\longrightarrow~&\,4\,\cV^{\cM}{}_{IK}\cV^{\cN}{}_{A}\left( \Gamma^{I K J}_{A \dot{A}} +12\,\Gamma^I_{A \dot{A}} \delta^{K J}\right)\bar\chi^{\dot{A}}\left\{\nabla_{\cM},\nabla_{\cN}\right\}\epsilon^J\nonumber\\
&+4\,\cV^{\cM}{}_{IK}\cV^{\cN}{}_{A}( \Gamma^{I K J}_{A \dot{A}} -10\Gamma^I_{A \dot{A}} \delta^{K J})\bar\chi^{\dot{A}}\left[\nabla_{\cM},\nabla_{\cN}\right]\epsilon^J\nonumber\\
&+16\,\cV^{\cN}{}_{IJ}\cV^{\cM}{}_{A}\bar\chi^{\dot{A}}\Gamma^{I}_{A\dot{A}}\,\gamma^{\mu}\nabla_{\cM}\nabla_{\cN}\gamma_{\mu} \,\epsilon^J\;.
\end{align}
Using the identity $\eqref{comintder2}$ and the section constraints (\ref{sectionso16}) 
one finds that all quadratic and linear terms in $\epsilon$ vanish while the remaining terms 
precisely cancel the last line of $\eqref{varpot}$.  For this, the following relations are useful
\bea
\nabla_{\cM}\gamma^{\mu\nu}&=&2\,\gamma^{[\mu}\nabla_{\cM}\gamma^{\nu]}\;, \qquad \gamma^{\mu}\nabla_{\cM}\gamma_{\mu}~=~0\;,\\
\gamma^{\nu}\nabla_{\cM}\nabla_{\cN}\gamma_{\nu}&=&-\frac12\,\cR_{\cM \cN}{}^{a b}\gamma_{a b}-\frac14\,\gamma^{\mu}\nabla_{\cM}g^{\nu \rho}\nabla_{\cN}g_{\nu \rho}\;.
\eea
We have thus sketched the vanishing of all terms carrying two internal derivatives in the supersymmetry
variation of (\ref{completelag}). 
The cancellation of the remaining terms is described in detail in appendix~\ref{app:susy}.
To summarize the result, we have shown invariance of the action (\ref{completelag}) 
up to quartic fermion terms.

%%%%%%%%%%%%%%%%%%%%%%%%%%%%%%%%
\section{Conclusions}
%%%%%%%%%%%%%%%%%%%%%%%%%%%%%%%%

In this paper we have constructed the supersymmetric completion of the bosonic ${\rm E}_{8(8)}$ exceptional field theory.
The final result is given by the action (\ref{completelag}) and the supersymmetry transformation laws (\ref{susyvar}).
In particular, we have established the supersymmetry algebra which consistently closes into the generalized
internal and external diffeomorphisms together with the tensor gauge transformations of the theory.
The geometry of the internal space is constrained by the section condition (\ref{section}) which admits (at least) two
inequivalent solutions for which the action (\ref{completelag}) reproduces the full $D=11$ supergravity
and full type IIB supergravity, respectively. The fermions of exceptional field theory can consistently accommodate 
the fermions of the type IIA and type IIB theory, since the ${\rm E}_{8(8)}$-covariant formulation (\ref{completelag}) 
does not preserve the original $D=10$ Lorentz invariance. The resulting $D=10$ fermion 
chirality thus depends on the solution of the section constraint.

In contrast to the standard formulation of supergravities, in exceptional field theory the bosonic symmetries already
uniquely determine the bosonic Lagrangian without any reference to fermions and supersymmetry. Nevertheless,
it is important to establish that the resulting bosonic Lagrangian allows for a supersymmetric completion upon coupling
of the proper fermionic field content as we have done in this paper. In particular, in the context of generalized Scherk-Schwarz 
reductions~\cite{Hohm:2014qga} this construction provides the consistent reduction formulas for the embedding of the
fermionic sector of lower-dimensional supergravities into higher dimensions.

A particular attribute of E$_{8(8)}$ exceptional field theory is the appearance of an additional
constrained gauge connection $B_{\mu\,\cM}$ related to an additional gauge symmetry which ensures
closure of the algebra of generalized diffeomorphisms. Unlike all other fields of E$_{8(8)}$ exceptional field theory,
this gauge connection is invisible in three-dimensional supergravity. More precisely, upon a consistent truncation 
of exceptional field theory down to three dimensions by means of a generalized Scherk-Schwarz reduction
\bea
{\cal M}_{\cM\cN}(x,Y) &=& U_{\cM}{}^{\cK}(Y)\,U_{\cN}{}^{\cL}(Y)\,M_{\cK\cL}(x)\;, 
\nonumber\\
 g_{\mu\nu}(x,Y) &=& \rho^{-2}(Y)\,g_{\mu\nu}(x)\;,\nonumber\\
  {A}_{\mu}{}^{\cM}(x,Y) &=& \rho^{-1}(Y) A_{\mu}{}^{\cN}(x)(U^{-1})_{\cN}{}^{\cM}(Y) \;, 
\eea 
with the $Y$-dependence carried by an E$_{8(8)}$ matrix $U$ and a scaling factor $\rho$
(satisfying their system of consistency equations), the constrained gauge connection $B_{\mu\,\cM}$
reduces according to
\bea
{B}_{\mu\,\cM}(x,Y) &\propto&
 \rho^{-1}(Y)\,(U^{-1})_\cK{}^\cP(Y) \,\partial_\cM U_\cP{}^\cL(Y) \,f_{\cN\cL}{}^{\cK}\, A_{\mu}{}^{\cN}(x) 
\;,
\eea
such that its fluctuations are expressed in terms of the same three-dimensional vector fields $A_{\mu}{}^{\cN}(x)$
that parametrize the fluctuations of the ${A}_{\mu}{}^{\cM}(x,Y)$\,.
It is thus tempting to wonder if already in exceptional field theory, and before reduction, 
the constrained gauge connection can be considered as a function of the remaining fields such as \cite{Cederwall:2015ica}
\bea
B_{\mu\,\cM}&\stackrel{?}=&
\Gamma_{\cM,\cN}\,A_\mu{}^\cN
\;,
\label{BA?}
\eea
c.f.\ (\ref{SigSig}). However, as seen above, coupling to fermions requires a connection $\Gamma_{\cM,\cN}$
other than the Weitzenb\"ock connection, such that (\ref{BA?}) would obstruct compatibility with the constraints (\ref{sectionB}).
Moreover, supersymmetry of the Lagrangian requires a non-trivial transformation law (\ref{susyvar}) for the constrained
connection $B_{\mu\,\cM}$.
It is remarkable that as we have shown above
this additional constrained connection consistently joins the remaining bosonic and fermionic
fields into a single supermultiplet without the need of additional fermionic matter.

The fact that all transformation laws of $B_{\mu\,\cM}$ are most compactly expressed 
in terms of the general covariant variation (\ref{DeltaB}) is remnant of structures that appear in 
a general tensor hierarchy of non-abelian $p$-forms \cite{deWit:2008ta}.
This may hint at a yet larger algebraic structure which in particular unifies the topological
term and the generalized three-dimensional Einstein-Hilbert term of (\ref{action_bos}) 
into a single non-abelian Chern-Simons form on an enlarged algebra.
If the present construction should allow for a generalization to 
the infinite-dimensional cases of E$_9$ \cite{Julia:1981wc,Julia:1982gx,Nicolai:1988jb},
E$_{10}$ \cite{Damour:2002cu,Damour:2006xu}, (and maybe E$_{11}$ \cite{West:2001as,West:2003fc}), 
this appearance of additional bosonic representations and their interplay with supersymmetry 
may play an essential role.

\section*{Acknowledgments}

We would like to thank Olaf Hohm and Ergin Sezgin for interesting and useful discussions.
The algebraic calculations in this paper have been facilitated in part by using the 
computer algebra system {\tt Cadabra} \cite{Peeters:2006kp,Peeters:2007wn}.

\newpage

\section*{Appendices}

\begin{appendix}
%%%%%%%%%%%%%%%%%%%%%%%%%%%%%%%%
% renumbering of equations
%%%%%%%%%%%%%%%%%%%%%%%%%%%%%%%%

\makeatletter
\@addtoreset{equation}{section}
\makeatother

\renewcommand{\thetable}{\Roman{table}}
\renewcommand{\theequation}{\Alph{section}.\arabic{equation}}
\renewcommand{\thesection}{\Alph{section}}

%%%%%%%%%%%%%%%%%%%%%%%%%%%%%%%%
\pagenumbering{roman}
\section{${\rm E_{8(8)}}$ conventions}
\label{app:conventions}

The ${\rm E_{8(8)}}$ generators $t^\cM$ split into 120 compact ones
$X^{IJ}\equiv -X^{JI}$ and 128 non-compact ones $Y^A$, with $SO(16)$
vector indices $I, J, \dots \in \mathbf{16}$\,, spinor indices $A,
\in \mathbf{128}$, and the collective label $\cM= ([IJ],A)$. The
conjugate $SO(16)$ spinors are labeled by dotted indices $\dot{A}, \dot{B},
\dots$.  In this $SO(16)$ basis the totally antisymmetric $E_{8(8)}$
structure constants $f^{\cM\cN\cK}$ possess the non-vanishing
components:
\begin{equation}\label{strcst}
f^{I\!J,\,K\!L,\,M\!N} = 
-8\, \delta\!\oversym{^{I[K}_{\vphantom{M}}\,\delta_{MN}^{L]J}}
\;,\qquad
f^{I\!J,\,A,\,B}   = -\frac{1}{2} {\Gamma}^{IJ}_{AB} \; .
\end{equation}
$E_{8(8)}$ indices are raised and lowered by means of the
Cartan-Killing metric
\begin{equation}\label{cartanmetric}
\eta^{\cM\cN}=\frac1{60} {\rm Tr} \, t^\cM t^\cN 
             =-\frac1{60} f^\cM{}_{\cK\cL}f^{\cN\cK\cL} \;,
\end{equation}
with components $\eta^{AB}=\delta^{AB}$ and
$\eta^{I\!J\,K\!L}=-2\delta^{IJ}_{KL}$.  When summing over
antisymmetrized index pairs $[IJ]$, an extra factor of $\frac12$ is
always understood.

We will also need the projector onto the adjoint representation
\bea
\P^\cM{}_\cN{}^\cK{}_\cL&=&\frac{1}{60}\,{f^{\cM}}_{\cN\cP}{f^{\cP\cK}}_\cL
\nonumber\\
&=&\frac{1}{30}\,\delta^{\cM}_{(\cN}\delta^{\cK}_{\cL)}+\frac{7}{30}\,(\P_{\mathbf{3875}})_{\cN\cL}{}^{\cM\cK}-\frac{1}{240}\,\eta^{\cM\cK}\eta_{\cN\cL}+\frac{1}{120}\,f^{\cM\cK}{}_\cP f^\cP{}_{\cN\cL}\;,
\label{Padj}
\eea
in terms of the Cartan-Killing form and structure constants of ${\rm E}_{8(8)}$ and
the projector $(\P_{\mathbf{3875}})_{\cN\cL}{}^{\cM\cK}$ explicitly given by
  \be\label{3875proj}
   (\mathbb{P}_{3875}){}^{{\cal MK}}{}_{{\cal NL}} \ = \ \frac{1}{7}\,\delta^{{\cal M}}_{({\cal N}}\, \delta^{{\cal K}}_{{\cal L})}
   -\frac{1}{56}\,\eta^{{\cal MK}}\,\eta_{{\cal NL}}-\frac{1}{14}\,f^{{\cal P}}{}_{{\cal N}}{}^{({\cal M}}\, f_{{\cal PL}}{}^{{\cal K})}\;. 
  \ee 
We refer to \cite{Koepsell:1999uj,Koepsell:2000xg} for other useful E$_{8(8)}$ identities.

\section{Gamma matrix identities}\label{gamid} 
In this appendix, we give some of the SO(16) gamma matrices identities we have used to rewrite the curvature $\cR_A$ in a more compact form. We started with 14 terms quadratic in the Cartan forms, where a simple counting gives only 12 independent terms. Then using an explicit representation of the $SO(16)$ gamma matrices together with the section constraints (\ref{sectionso16}), we were able to write 
$\cR_A$ with 7 independent terms quadratic in the Cartan forms.

The main identities behind this simplification are the following
\bea
\cV^{[\cM}{}_{IJ}\cV^{\cN]}{}_B(\Gamma^{IJ}\Gamma^{KL})_{BD}p_{\cM}{}^A p_{\cN}{}^C&=&0\,,\\
\Gamma^{IM}_{B[A}\Gamma^{IMNP}_{D]C}\cV^{\cM}{}_{NP}\cV^{\cN}{}_{B}p_{\cM}{}^C p_{\cN}{}^D&=&-4\Gamma^{IM}_{A[B}\Gamma^{IN}_{C]D}\cV^{\cM}{}_{MN}\cV^{\cN}{}_{B}p_{\cM}{}^Cp_{\cN}{}^D\nonumber\\
&&+8\Gamma^{IM}_{AB}\cV^{\cM}{}_{IM}\cV^{\cN}{}_{C}p_{\cM}{}^{(B}p_{\cN}{}^{C)}\nonumber\\
&&-\Gamma^{IM}_{BC}\cV^{\cM}{}_{IM}\cV^{\cN}{}_{B}p_{\cM}{}^Ap_{\cN}{}^C\nonumber\\
&&-\Gamma^{IM}_{AB}\cV^{\cM}{}_{IM}\cV^{\cN}{}_{B}p_{\cM}{}^Cp_{\cN}{}^C\;.
\eea

\section{Supersymmetry of the Lagrangian}
\label{app:susy}

In this appendix, we give the remaining details for the invariance of the Lagrangian (\ref{completelag}) 
under the supersymmetry transformations $\eqref{susyvar}$.

\subsection{Cancellation of the terms carrying field strengths}

We start with a simple check: all terms in $\cF_{\mu\nu}{}^{\cM}$ and $\cG_{\mu\nu\,\cM}$ from the supersymmetric variation of the fermionic terms in the Lagrangian should cancel against the corresponding contributions from
variation of the kinetic and topological terms. The relevant contribution on the fermionic side are
\bea
\delta \left( -2\,e \bar\chi^{\dot{A}} \gamma^\mu\gamma^\nu \psi_\mu^I \,\Gamma^I_{A\dot{A}}\,{\cal P}_\nu^A \right)
&\longrightarrow&
2\,e \bar\chi^{\dot{A}} \gamma^{\mu\nu}  \epsilon^I \,\Gamma^I_{A\dot{A}}\,{\cal D}_\mu {\cal P}_\nu^A
 \nonumber\\
 &=&
-i  \varepsilon^{\mu\nu\rho}  \bar\chi^{\dot{A}} \gamma_{\rho}  \epsilon^I \,\Gamma^I_{A\dot{A}}
 {\cal V}^{\cM A}
 \left(
\widetilde{\cG}_{\mu\nu\,\cM}
-f_{\cM\cL}{}^{\cK} \nabla_{\cK}\cF_{\mu\nu}{}^{\cL}
\right)
\nonumber\\
\delta\left(
2\,\varepsilon^{\mu\nu\rho} \bar\psi_\mu^I {\cal D}_{\nu} \psi_\rho^I
\right) &\longrightarrow&
2\,\varepsilon^{\mu\nu\rho} \bar\psi_\mu^I \left[{\cal D}_{\nu}\,, {\cal D}_\rho\right] \epsilon^I
\nonumber\\
&\longrightarrow&
+2\,\varepsilon^{\mu\nu\rho} \,\cV^{\cM}{}_{IJ}
\left(
\widetilde{\cG}_{\mu\nu\,\cM}
-f_{\cM\cL}{}^{\cK} \nabla_{\cK}\cF_{\mu\nu}{}^{\cL}
\right)
\bar\psi_\rho^I\epsilon^J
\nonumber\\
&&{}
+\varepsilon^{\mu\nu\rho} \,\cF_{\mu\nu}{}^\cM\left(
 \nabla_{\cM} \bar\psi_\rho^I\epsilon^I - \bar\psi_\rho^I \nabla_{\cM}\epsilon^I \right)
 \;,
\eea
where we have used the commutator of two external covariant derivative $\eqref{comderextcov}$.
On the bosonic side, all terms with field strength come from the variation of kinetic and topological terms
\bea
\delta\cL&\longrightarrow&
\varepsilon^{\mu\nu\rho}\tilde{\cal E}_{\mu\nu\,\cM}^{(B)}\,\delta A_\rho{}^{\cM}
+\varepsilon^{\mu\nu\rho}{\cal E}_{\mu\nu}^{(A)\,\cM}\,\Delta B_{\rho \cM}
\nonumber\\
&\longrightarrow&
\varepsilon^{\mu\nu\rho}(-\frac12\,\widetilde{\cG}_{\mu\nu\,{\cal M}} 
+\frac12\,f_{\cM \cN}{}^\cK\nabla_\cK{\cal F}_{\mu\nu}{}^{\cN})(-4\,
\cV^{\cM}{}_{IJ}{\bar{\epsilon}}^I{\psi_{\rho}}^J
+2i\,
\Gamma{}^I_{A\dot{A}}\cV^{\cM}{}_A{\bar{\epsilon}}^I \gamma_{\mu}\chi^{\dot{A}})
\nonumber\\
&&{}
\varepsilon^{\mu\nu\rho}{\cal F}_{\mu\nu}{}^{\cM}\,
\left(\nabla_{\cM}\bar{\epsilon}^I \psi_{\rho}{}^I-\bar{\epsilon}^I \nabla_{\cM} \psi_{\rho}{}^I\right)
+{\cal F}_{\mu\nu}{}^{\cM}\,
g^{\sigma \mu}\nabla_\cM(\bar{\epsilon}^Ii\gamma^{\nu}\psi_{\sigma}{}^I)
\;,
\label{varvectf}
\eea
with the exception of an extra contribution from the improved Einstein-Hilbert term
\bea
\delta\left(-e e_a{}^\mu e_b{}^\nu {\cal F}_{\mu\nu}{}^{\cM}\,\omega_\cM{}^{ab}\right)
&\longrightarrow&
-e e_a{}^\mu e_b{}^\nu {\cal F}_{\mu\nu}{}^{\cM}\,
\delta \omega_\cM{}^{ab}
\nonumber\\
&=&
-e e_a{}^\mu e_b{}^\nu {\cal F}_{\mu\nu}{}^{\cM}\left(
\delta e^{\rho[a} \nabla_{\cM} e_\rho{}^{b]}+ e^{\rho[a} \nabla_{\cM} \delta e_\rho{}^{b]}
\right)
\nonumber\\
&=&
-e e_a{}^\mu e_b{}^\nu {\cal F}_{\mu\nu}{}^{\cM}\left(
e^\sigma{}^{[a} e_{\tau}{}^{b]}\,\nabla_\cM\left(e_c{}^{\tau} \delta e_\sigma{}^c\right)
\right)
\nonumber\\
&=&
-i e g^{\mu\sigma}  {\cal F}_{\mu\nu}{}^{\cM}
 \nabla_\cM\left(\bar\epsilon^I \gamma^\nu \psi_\sigma^I  \right)
\;.
\eea
that cancels the last term of $\eqref{varvectf}$.
Together, all terms with field strengths vanish.

\subsection{Cancellation of the $\nabla_M {\cal D}_\mu \chi \epsilon$ terms}

From the variation of the vector fields in the bosonic Lagrangian (we have now dropped all terms with field strengths),
we have the following contribution
\bea
\delta\cL&\longrightarrow& 
+e j^{\mu\,\cM}\,\Delta B_{\mu\, \cM}
- e\,f_{\cM \cN}{}^\cK\nabla_\cK j^{\mu\,\cN}\,\delta A_\mu{}^{\cM}
-e\,\widehat{J}^\mu{}_{\cM} \,\delta A_\mu{}^{\cM} 
\nonumber\\
&\longrightarrow&
+4ie\,f_{\cM \cN}{}^\cK\nabla_\cK ({\cal V}^\cN{}_B {\cal P}^{\mu\,B})\,{\cal V}^\cM{}_A \Gamma^I_{A\dot{A}} 
\bar\chi^{\dot{A}}\gamma_\mu \epsilon^I
+2i e\,\widehat{J}^\mu{}_{\cM} \,{\cal V}^\cM{}_A \Gamma^I_{A\dot{A}} 
\bar\chi^{\dot{A}}\gamma_\mu \epsilon^I
\nonumber\\
&=&
- ie\, {\cal V}^\cM{}_{KL}  \nabla_\cM {\cal P}^{\mu\,A}\, \Gamma^{IKL}_{A\dot{A}} 
\bar\chi^{\dot{A}}\gamma_\mu \epsilon^I
 -2ie\, {\cal V}^\cM{}_{IJ}  \nabla_\cM {\cal P}^{\mu\,A}\, \Gamma^I_{A\dot{A}} 
\bar\chi^{\dot{A}}\gamma_\mu \epsilon^J
\nonumber\\
&&{}
+2i e\,\widehat{J}^\mu{}_{\cM} \,{\cal V}^\cM{}_A \Gamma^I_{A\dot{A}} 
\bar\chi^{\dot{A}}\gamma_\mu \epsilon^I
\;.
\eea
On the fermionic side, the relevant contributions to this sector are
\bea
\delta\left(
 -2\,e \bar\chi^{\dot{A}} \gamma^\mu\gamma^\nu \psi_\mu^I \,\Gamma^I_{A\dot{A}}\,{\cal P}_\nu^A 
\right)
&\longrightarrow&
 -4\,i\,e \bar\chi^{\dot{A}} \gamma^\mu\gamma^\nu 
 \nabla_\cM(\gamma_\mu \epsilon ^J) \,\Gamma^I_{A\dot{A}}\,{\cal P}_\nu^A {\cal V}^\cM{}_{IJ}
\nonumber\\
&&{}
 +4\,i\,e \bar\chi^{\dot{A}} \gamma^\mu \nabla_\cM \epsilon ^J \,\Gamma^I_{A\dot{A}}\,{\cal P}_\mu^A {\cal V}^\cM{}_{IJ}\;,\\ 
\delta
\left(
-2i\,e\,\bar\chi^{\dot{A}} \gamma^\mu {\cal D}_\mu \chi^{\dot{A}}
\right)
&\longrightarrow&
8i\,e\,\bar\chi^{\dot{A}} \gamma^\mu {\cal D}_\mu \nabla_{\cM} \epsilon^I {\cal V}^{\cM}{}_A \Gamma^I_{A\dot{A}}
\nonumber\\
&&{}
-2i\,e\,\bar\chi^{\dot{A}} \gamma^\mu  \nabla_{\cM} \epsilon^I  {\cal P}_\mu^A
 {\cal V}^{\cM}{}_{JK} \Gamma^{IJK}_{A\dot{A}}
\nonumber\\
&&{}
-4i\,e\,\bar\chi^{\dot{A}} \gamma^\mu  \nabla_{\cM} \epsilon^J  {\cal P}_\mu^A 
 {\cal V}^{\cM}{}_{IJ} \Gamma^I_{A\dot{A}}\;,\\
\delta\left(
-8\,e\, \cV^{\cM}{}_{A}\Gamma^I_{A\dot{A}}\bar{\psi}_{\mu}{}^Ii\nabla_{\cM}(\gamma^{\mu}\chi^{\dot{A}})
\right) 
&\longrightarrow&
8\,i\, e\,\cV^{\cM}{}_{A}\Gamma^I_{A\dot{A}}
\nabla_{\cM}\bar\chi^{\dot{A}} \gamma^{\mu}
{\cal D}_\mu {\epsilon}^I
\nonumber\\
&=&
-8\,i\, e\,\cV^{\cM}{}_{A}\Gamma^I_{A\dot{A}}
\bar\chi^{\dot{A}} \gamma^{\mu}
\nabla_{\cM} {\cal D}_\mu {\epsilon}^I\;,
\\
\delta\left(
-2 e\,\cV^{\cM}{}_{IJ}\Gamma^{IJ}_{\dot{A}\dot{B}}\bar{\chi}^{\dot{A}}\nabla_{\cM}\chi^{\dot{B}}
\right) 
&\longrightarrow&
-2i\,e\, \cV^{\cM}{}_{IJ}\Gamma^{IJ}_{\dot{A}\dot{B}}\bar{\chi}^{\dot{A}}\nabla_{\cM}(\gamma^\mu \epsilon^K
\Gamma^K_{A\dot{B}} {\cal P}_\mu^A)
\nonumber\\
&=&
2i\,e\, \cV^{\cM}{}_{IJ}\Gamma^{IJK}_{A\dot{A}}\bar{\chi}^{\dot{A}}\nabla_{\cM}(\gamma^\mu \epsilon^K
 {\cal P}_\mu^A)
 \nonumber\\
 &&{}
-4i\,e\, \cV^{\cM}{}_{IJ}\Gamma^{I}_{A\dot{A}}\bar{\chi}^{\dot{A}}\nabla_{\cM}(\gamma^\mu \epsilon^J
 {\cal P}_\mu^A)
\;.
\eea
Using the commutator
\bea
\cV^{\cM}{}_A\Gamma^I_{A\dot{A}}
\,
\left[\nabla_{\cM}, {\cal D}_\mu \right]{\epsilon}^I
&=&
\frac14\,\cV^{\cM}{}_{A}\Gamma^I_{A\dot{A}}
\,
{\cal R}_{\cM \mu}{}^{ab}\,\gamma_{ab} {\epsilon}^I
\nonumber\\
&&{}
-\frac34\,\Gamma^{I}_{A\dot{A}} {\cal V}^\cM{}_{IJ} \nabla_\cM {\cal P}_\mu{}^A \,\epsilon^J
+\frac18\,\Gamma^{IJK}_{A\dot{A}} {\cal V}^\cM{}_{IJ} \nabla_\cM {\cal P}_\mu{}^A \,\epsilon^K\;,
\eea
all of the above terms simply reduce to
\bea
&\longrightarrow&
 -4\,i\,e \bar\chi^{\dot{A}} \gamma^\mu\gamma^\nu 
 \nabla_\cM (\gamma_\mu) \epsilon ^J \,\Gamma^I_{A\dot{A}}\,{\cal P}_\nu^A {\cal V}^\cM{}_{IJ}
 -4i\,e\, \cV^{\cM}{}_{IJ}\Gamma^{I}_{A\dot{A}}\bar{\chi}^{\dot{A}}\nabla_{\cM}(\gamma^\mu) \epsilon^J
 {\cal P}_\mu^A
\nonumber\\
&&{}
 -2ie\, {\cal V}^\cM{}_{IJ}  (\nabla_\cM g^{\mu\nu}){\cal P}_\nu^{A}\, \Gamma^I_{A\dot{A}} 
\bar\chi^{\dot{A}}\gamma_\mu \epsilon^J
\nonumber\\
&&{}
-2i\,e\,\cV^{\cM}{}_{A}\Gamma^I_{A\dot{A}}
\,
{\cal R}_{\cM \mu}{}^{ab}\,\bar\chi^{\dot{A}} \gamma^{\mu} \gamma_{ab} {\epsilon}^I
+2i e\,\widehat{J}^\mu{}_{\cM} \,{\cal V}^\cM{}_A \Gamma^I_{A\dot{A}} 
\bar\chi^{\dot{A}}\gamma_\mu \epsilon^I
\nonumber\\
&=&
 4i\,e\,{\cal V}^\cM{}_A\Gamma^I_{A\dot{A}}
\left(
{\cal R}_{\cM \nu}{}^{\mu \nu}\,
+\frac12\,\widehat{J}^\mu{}_{\cM}\right)
\bar\chi^{\dot{A}} \gamma_{\mu} {\epsilon}^I\nonumber\\
&=& 0\;,
\eea
where we have used $\eqref{BIR2}$ in the last equality.

\subsection{Cancellation of the $\nabla_M {\cal D}_\mu \psi \epsilon$ terms}

Similarly, we collect the vector field contributions in the bosonic Lagrangian
\bea
\delta\cL&\longrightarrow& 
+e j^{\mu\,\cM}\,\Delta B_{\mu\, \cM}
- e\,f_{\cM \cN}{}^\cK\nabla_\cK j^{\mu\,\cN}\,\delta A_\mu{}^{\cM}
-e\,\widehat{J}^\mu{}_{\cM} \,\delta A_\mu{}^{\cM} 
\nonumber\\
&\longrightarrow& 2 e \cP^{\mu}{}^A\cV^{\cM}{}_A\left(-4 \bar{\psi}_{\mu}{}^I\nabla_{\cM}\epsilon^I+2\nabla_{\cM}(\bar{\psi}_{\mu}{}^I\epsilon^I)
+e\,\varepsilon_{\mu\nu\rho}g^{\rho\sigma}\nabla_\cM(\bar{\epsilon}^Ii\gamma^{\nu}\psi_{\sigma}{}^I)\right)
\nonumber\\
&& -8e\,\cV^{\cN}{}_A\cV^\cM{}_{IJ}\,f_{\cM \cN}{}^\cK\nabla_\cK \cP^{\mu}{}^A\bar{\psi}^I_{\mu}\epsilon^J-4e\widehat{J}^\mu{}_{\cM}\cV^\cM{}_{IJ}\bar{\psi}^I_{\mu}\epsilon^J
\nonumber\\
&=&-8e \cV^{\cM}{}_A\cP^{\mu}{}^A\bar{\psi}^I_{\mu}\nabla_{\cM}\epsilon^I-4e \cV^{\cM}{}_A\nabla_{\cM}(g^{\mu\nu}\cP_{\nu}{}^A)\bar{\psi}^I_{\mu}\epsilon^I+2\varepsilon^{\mu\nu\rho} \cV^{\cM}{}_A\nabla_\cM(\cP_{\mu}{}^A)\bar{\psi}_{\rho}{}^Ii\gamma^{\nu}\epsilon^I
\nonumber\\
&&+2\varepsilon^{\mu\lambda\rho} \cV^{\cM}{}_A\cP_{\mu}{}^A\nabla_\cM(g_{\lambda \nu})\bar{\psi}_{\rho}{}^Ii\gamma^{\nu}\epsilon^I
\nonumber\\
&&-4e\,\Gamma^{I J}_{AB}\cV^\cM{}_A\nabla_\cM \cP^{\mu}{}^B\bar{\psi}^I_{\mu}\epsilon^J-4e\widehat{J}^\mu{}_{\cM}\cV^\cM{}_{IJ}\bar{\psi}^I_{\mu}\epsilon^J\;,
\eea 
together with the relevant contributions from the fermionic Lagragian
\bea
\delta(2ie\gamma^{\mu\nu\rho}\bar{\psi}^I_{\rho}\cD_{\mu}\psi_{\nu}^I)
&\longrightarrow&
8 ie \gamma^{\mu\nu\rho} \cV^\cM{}_{IJ}\bar{\psi}^I_{\rho} \cD_{\mu}(\nabla_{\cM}(i\gamma_{\nu}\epsilon^J)+i\gamma_{\nu}\nabla_{\cM}\epsilon^J)
\nonumber\\
&&-4 ie \gamma^{\mu\nu\rho} \bar{\psi}^I_{\rho}\cP_{\mu}{}^A \Gamma^{IJ}_{AB} \cV^{\cM}{}_B(\nabla_{\cM}(i\gamma_{\nu}\epsilon^J)+i\gamma_{\nu}\nabla_{\cM}\epsilon^J)
\nonumber\\
&=&-16 e \cV^\cM{}_{IJ}\bar{\psi}^I_{\mu}\gamma^{\mu\nu}  \cD_{\nu}\nabla_{\cM}\epsilon^J
+8e \bar{\psi}_{\mu}^I\gamma^{\mu\nu}\nabla_{\cM}\epsilon^J\cP_{\mu}{}^A\Gamma^{IJ}_{AB}\cV^{\cM}{}_B
\nonumber\\
&&+8 ie \varepsilon^{\mu\nu\rho} \cV^\cM{}_{IJ}\bar{\psi}^I_{\mu} \cD_{\nu}(\nabla_{\cM}(\gamma_{\rho})\epsilon^J)
\nonumber\\
&&-4ie \varepsilon^{\mu\nu\rho} \bar{\psi}_{\mu}^I\nabla_{\cM}\gamma_{\rho}\epsilon^J\cP_{\mu}{}^A\Gamma^{IJ}_{AB}\cV^{\cM}{}_B\;,\\
\delta(-2e\bar\chi^{\dot{A}}\gamma^{\mu}\gamma^{\nu}\psi^I_{\mu}\Gamma^I_{A\dot{A}}\cP_{\nu}{}^A)
&\longrightarrow&
4e \cV^\cM{}_A (\Gamma^J\Gamma^I)_{AB}\cP_{\nu}{}^B \nabla_{\cM}\bar{\epsilon}^J\gamma^{\mu}\gamma^{\nu}\psi^I_{\mu}
\nonumber\\
&=& -4e \cV^\cM{}_A (\Gamma^J\Gamma^I)_{AB}\cP_{\nu}{}^B \bar{\psi}^I_{\mu}\gamma^{\mu \nu}\nabla_{\cM}\epsilon^J
\nonumber\\
&& +4e \cV^\cM{}_A (\Gamma^J\Gamma^I)_{AB}\cP^{\mu}{}^B \bar{\psi}^I_{\mu}\nabla_{\cM}\epsilon^J\;,\\
\delta(8e\cV^{\cM}{}_{IJ}\bar{\psi}_{\mu}{}^I\gamma^{\mu\nu}\nabla_{\cM}\psi_{\nu}{}^J)
&\longrightarrow&
-8ie\varepsilon^{\mu\nu\rho}\cV^{\cM}{}_{IJ}\bar{\psi}_{\mu}{}^I\nabla_{\cM}\gamma_{\rho}\cD_{\nu}\epsilon^J
\nonumber\\
&&+16e\cV^{\cM}{}_{IJ}\bar{\psi}_{\mu}{}^I\gamma^{\mu\nu}\nabla_{\cM}\cD_{\nu}\epsilon^J
\nonumber\\
&=&-8ie\varepsilon^{\mu\nu\rho}\cV^{\cM}{}_{IJ}\bar{\psi}_{\mu}{}^I\cD_{\nu}(\nabla_{\cM}\gamma_{\rho}\epsilon^J)
\nonumber\\
&&+8ie\varepsilon^{\mu\nu\rho}\cV^{\cM}{}_{IJ}\bar{\psi}_{\mu}{}^I\cD_{\nu}(\nabla_{\cM}\gamma_{\rho})\epsilon^J
\nonumber\\
&&+16e\cV^{\cM}{}_{IJ}\bar{\psi}_{\mu}{}^I\gamma^{\mu\nu}\nabla_{\cM}\cD_{\nu}\epsilon^J\;,\\
\delta(-8ie\bar{\psi}_{\mu}{}^I\nabla_{\cM}(\gamma^{\mu}\chi^{\dot{A}})\Gamma^I_{A\dot{A}}\cV^\cM{}_A)
&\longrightarrow&
4 e \cV^{\cM}{}_A(\Gamma^{I}\Gamma^J)_{AB} \bar{\psi}_{\mu}^I\nabla_{\cM}(\gamma^{\mu}\gamma^{\nu}\epsilon^J\cP_{\nu}{}^B)
\nonumber\\
&=& 4 e \cV^{\cM}{}_A(\Gamma^{I}\Gamma^J)_{AB} \bar{\psi}_{\mu}^I\nabla_{\cM}(\gamma^{\mu}\gamma^{\nu})\epsilon^J\cP_{\nu}{}^B
\nonumber\\
&& +4 e \cV^{\cM}{}_A(\Gamma^{I}\Gamma^J)_{AB} \bar{\psi}_{\mu}^I\gamma^{\mu \nu}\nabla_{\cM}\epsilon^J\cP_{\nu}{}^B
\nonumber\\
&& + 4 e \cV^{\cM}{}_A(\Gamma^{I}\Gamma^J)_{AB} \bar{\psi}_{\mu}^I\nabla_{\cM}\epsilon^J\cP^{\mu}{}^B
\nonumber\\
&& + 4 e \cV^{\cM}{}_A(\Gamma^{I}\Gamma^J)_{AB} \bar{\psi}_{\mu}^I\gamma^{\mu}\gamma^{\nu}\epsilon^J\nabla_{\cM}\cP_{\nu}{}^B\;.
\eea
Upon using the commutator
\bea
\cV^{\cM}{}_{IJ}
\,
\left[\nabla_{\cM}, {\cal D}_\mu \right]{\epsilon}^J
&=&
\frac14\,\cV^{\cM}{}_{IJ}
\,
{\cal R}_{\cM \mu}{}^{ab}\,\gamma_{ab} {\epsilon}^J
\nonumber\\
&&{}
-\frac18\,{\cal V}^\cM{}_A \nabla_\cM {\cal P}_\mu{}^A \,\epsilon^I
-\frac14\,\Gamma^{IJ}_{AB} {\cal V}^\cM{}_A \nabla_\cM {\cal P}_\mu{}^A \,\epsilon^J\;,
\eea
this reduces to
\bea
&\longrightarrow& -8 e \cV^{\cM}{}_{IJ}\bar\psi_{\mu}^I\epsilon^J(\cR_{\cM \nu}{}^{\mu\nu}+\frac12\widehat{J}^\mu{}_{\cM})\nonumber\\
&&+8i e\varepsilon^{\nu\rho\sigma} \cV^{\cM}{}_{IJ}\bar\psi_{\mu}^I\gamma_{\sigma}\epsilon^J\cR_{\cM\nu \kappa\rho}g^{\kappa\mu}
\nonumber\\
&&-8i e\varepsilon^{\nu\rho\sigma} \cV^{\cM}{}_{IJ}\bar\psi_{\nu}^I\gamma_{\sigma}\epsilon^J\cR_{\cM\mu \kappa\sigma}g^{\kappa\mu}
\nonumber\\
&&-8ie\varepsilon^{\mu\nu\rho}\cV^{\cM}{}_{IJ}\bar\psi_{\rho}^I([\cD_{\mu},\nabla_{\cM}]\gamma_{\nu})\epsilon^J
\nonumber\\
&=& 0
\;,
\eea
where we have used the Schouten identity
\bea
\varepsilon^{\nu\rho\sigma}g^{\kappa\mu}(\bar{\psi}_{\mu}^I\gamma_{\nu}\epsilon^J\cR_{\cM \rho \kappa \sigma}-\bar{\psi}_{\nu}^I\gamma_{\mu}\epsilon^J\cR_{\cM \rho \kappa \sigma}+\bar{\psi}_{\nu}^I\gamma_{\rho}\epsilon^J\cR_{\cM \mu \kappa \sigma})&=&\varepsilon^{\nu\rho\sigma}\bar{\psi}_{\nu}^I\gamma_{\rho}\epsilon^J\cR_{\cM \sigma \kappa \mu}g^{\kappa\mu}\;,\nonumber\\
&=&0\;.
\eea
This completes the results obtained in section 4 and 
proves the invariance of the extended Lagrangian $\eqref{completelag}$ under supersymmetry.
\end{appendix}

%\bibliographystyle{JHEP2}
%\bibliography{refs}

\begin{thebibliography}{10}

\bibitem{Hohm:2013pua}
O.~Hohm and H.~Samtleben, { Exceptional form of ${D}=11$ supergravity},  {
  Phys.Rev.Lett.} { 111} (2013) 231601,
[\href{http://xxx.lanl.gov/abs/1308.1673}{{\tt 1308.1673}}].
%%CITATION = ARXIV:1308.1673;%%.

\bibitem{Hohm:2013vpa}
O.~Hohm and H.~Samtleben, { Exceptional field theory {I}: ${E}_{6(6)}$
  covariant form of {M}-theory and type {IIB}},  { Phys.Rev.} { D89} (2014)
  066016,
[\href{http://xxx.lanl.gov/abs/1312.0614}{{\tt 1312.0614}}].
%%CITATION = ARXIV:1312.0614;%%.

\bibitem{Hohm:2013uia}
O.~Hohm and H.~Samtleben, { Exceptional field theory {II}: {E}$_{7(7)}$},  {
  Phys.Rev.} { D89} (2014) 066017,
[\href{http://xxx.lanl.gov/abs/1312.4542}{{\tt 1312.4542}}].
%%CITATION = ARXIV:1312.4542;%%.

\bibitem{Hohm:2014fxa}
O.~Hohm and H.~Samtleben, { Exceptional field theory {III}: ${E}_{8(8)}$},  {
  Phys.Rev.} { D90} (2014) 066002,
[\href{http://xxx.lanl.gov/abs/1406.3348}{{\tt 1406.3348}}].
%%CITATION = ARXIV:1406.3348;%%.

\bibitem{Cremmer:1978ds}
E.~Cremmer and B.~Julia, { The ${N}=8$ supergravity theory. 1. {T}he
  {L}agrangian},  { Phys.Lett.} { B80} (1978)
48.
%%CITATION = PHLTA,B80,48;%%.

\bibitem{Cremmer:1979up}
E.~Cremmer and B.~Julia, { The ${SO}(8)$ supergravity},  { Nucl. Phys.} { B159}
  (1979)
141.
%%CITATION = NUPHA,B159,141;%%.

\bibitem{Cremmer:1980gs}
E.~Cremmer, { Supergravities in 5 dimensions},  in { Superspace and
  supergravity : proceedings} (S.~Hawking and M.~Rocek., eds.), Cambridge Univ.
  Press, 1980.
\newblock Nuffield Gravity Workshop, Cambridge.

\bibitem{Coimbra:2011ky}
A.~Coimbra, C.~Strickland-Constable, and D.~Waldram, { {$E_{d(d)} \times
  \mathbb{R}^+$ generalised geometry, connections and M theory}},  { JHEP} {
  1402} (2014) 054,
[\href{http://xxx.lanl.gov/abs/1112.3989}{{\tt 1112.3989}}].
%%CITATION = ARXIV:1112.3989;%%.

\bibitem{Berman:2012vc}
D.~S. Berman, M.~Cederwall, A.~Kleinschmidt, and D.~C. Thompson, { {The gauge
  structure of generalised diffeomorphisms}},  { JHEP} { 1301} (2013) 064,
[\href{http://xxx.lanl.gov/abs/1208.5884}{{\tt 1208.5884}}].
%%CITATION = ARXIV:1208.5884;%%.

\bibitem{Berman:2010is}
D.~S. Berman and M.~J. Perry, { Generalized geometry and {M} theory},  { JHEP}
  { 1106} (2011) 074,
[\href{http://xxx.lanl.gov/abs/1008.1763}{{\tt 1008.1763}}].
%%CITATION = ARXIV:1008.1763;%%.

\bibitem{Berman:2011jh}
D.~S. Berman, H.~Godazgar, M.~J. Perry, and P.~West, { Duality invariant
  actions and generalised geometry},  { JHEP} { 1202} (2012) 108,
[\href{http://xxx.lanl.gov/abs/1111.0459}{{\tt 1111.0459}}].
%%CITATION = ARXIV:1111.0459;%%.

\bibitem{Hohm:2015xna}
O.~Hohm and Y.-N. Wang, { Tensor hierarchy and generalized {C}artan calculus in
  {SL}(3) $\times$ {SL}(2) exceptional field theory},  { JHEP} { 1504} (2015)
  050,
[\href{http://xxx.lanl.gov/abs/1501.01600}{{\tt 1501.01600}}].
%%CITATION = ARXIV:1501.01600;%%.

\bibitem{Abzalov:2015ega}
A.~Abzalov, I.~Bakhmatov, and E.~T. Musaev, { {Exceptional field theory:
  $SO(5,5)$}},  { JHEP} { 06} (2015) 088,
[\href{http://xxx.lanl.gov/abs/1504.01523}{{\tt 1504.01523}}].
%%CITATION = ARXIV:1504.01523;%%.

\bibitem{Musaev:2015ces}
E.~T. Musaev, { Exceptional field theory: ${SL}(5)$},  { JHEP} { 02} (2016)
  012,
[\href{http://xxx.lanl.gov/abs/1512.02163}{{\tt 1512.02163}}].
%%CITATION = ARXIV:1512.02163;%%.

\bibitem{Berman:2015rcc}
D.~S. Berman, C.~D.~A. Blair, E.~Malek, and F.~J. Rudolph, { An action for
  {F}-theory: $\mathrm{SL}(2) \times \mathbb{R}^+$ exceptional field theory},
\href{http://xxx.lanl.gov/abs/1512.06115}{{\tt 1512.06115}}.
%%CITATION = ARXIV:1512.06115;%%.

\bibitem{Godazgar:2014nqa}
H.~Godazgar, M.~Godazgar, O.~Hohm, H.~Nicolai, and H.~Samtleben, {
  Supersymmetric {E$_{7(7)}$} exceptional field theory},  { JHEP} { 1409}
  (2014) 044,
[\href{http://xxx.lanl.gov/abs/1406.3235}{{\tt 1406.3235}}].
%%CITATION = ARXIV:1406.3235;%%.

\bibitem{Musaev:2014lna}
E.~Musaev and H.~Samtleben, { Fermions and supersymmetry in {E}$_{6(6)}$
  exceptional field theory},  { JHEP} { 1503} (2015) 027,
[\href{http://xxx.lanl.gov/abs/1412.7286}{{\tt 1412.7286}}].
%%CITATION = ARXIV:1412.7286;%%.

\bibitem{Curtright:1980yk}
T.~Curtright, { Generalized gauge fields},  { Phys.Lett.} { B165} (1985)
304.
%%CITATION = PHLTA,B165,304;%%.

\bibitem{Hull:2000zn}
C.~Hull, { Strongly coupled gravity and duality},  { Nucl.Phys.} { B583} (2000)
  237--259,
[\href{http://xxx.lanl.gov/abs/hep-th/0004195}{{\tt hep-th/0004195}}].
%%CITATION = HEP-TH/0004195;%%.

\bibitem{West:2001as}
P.~C. West, { {${{E}}_{11}$ and {M} theory}},  { Class. Quant. Grav.} { 18}
  (2001) 4443--4460,
[\href{http://xxx.lanl.gov/abs/hep-th/0104081}{{\tt hep-th/0104081}}].
%%CITATION = HEP-TH/0104081;%%.

\bibitem{Hull:2001iu}
C.~Hull, { Duality in gravity and higher spin gauge fields},  { JHEP} { 0109}
  (2001) 027,
[\href{http://xxx.lanl.gov/abs/hep-th/0107149}{{\tt hep-th/0107149}}].
%%CITATION = HEP-TH/0107149;%%.

\bibitem{Cederwall:2015ica}
M.~Cederwall and J.~A. Rosabal, { {E}$_{8}$ geometry},  { JHEP} { 07} (2015)
  007,
[\href{http://xxx.lanl.gov/abs/1504.04843}{{\tt 1504.04843}}].
%%CITATION = ARXIV:1504.04843;%%.

\bibitem{Coimbra:2012af}
A.~Coimbra, C.~Strickland-Constable, and D.~Waldram, { Supergravity as
  generalised geometry {II}: {$E_{d(d)} \times \mathbb{R}^+$} and {M} theory},
  { JHEP} { 1403} (2014) 019,
[\href{http://xxx.lanl.gov/abs/1212.1586}{{\tt 1212.1586}}].
%%CITATION = ARXIV:1212.1586;%%.

\bibitem{Nicolai:1986jk}
H.~Nicolai, { ${D} = 11$ supergravity with local ${SO}(16)$ invariance},  {
  Phys. Lett.} { B187} (1987)
316.
%%CITATION = PHLTA,B187,316;%%.

\bibitem{Melosch:1997wm}
S.~Melosch and H.~Nicolai, { New canonical variables for $d = 11$
  supergravity},  { Phys. Lett.} { B416} (1998) 91--100,
[\href{http://xxx.lanl.gov/abs/hep-th/9709227}{{\tt hep-th/9709227}}].
%%CITATION = HEP-TH/9709227;%%.

\bibitem{Koepsell:2000xg}
K.~Koepsell, H.~Nicolai, and H.~Samtleben, { An exceptional geometry for {$D =
  11$} supergravity?},  { Class.Quant.Grav.} { 17} (2000) 3689--3702,
[\href{http://xxx.lanl.gov/abs/hep-th/0006034}{{\tt hep-th/0006034}}].
%%CITATION = HEP-TH/0006034;%%.

\bibitem{Marcus:1983hb}
N.~Marcus and J.~H. Schwarz, { Three-dimensional supergravity theories},  {
  Nucl. Phys.} { B228} (1983)
145.
%%CITATION = NUPHA,B228,145;%%.

\bibitem{Nicolai:2001sv}
H.~Nicolai and H.~Samtleben, { Compact and noncompact gauged maximal
  supergravities in three dimensions},  { JHEP} { 04} (2001) 022,
[\href{http://xxx.lanl.gov/abs/hep-th/0103032}{{\tt hep-th/0103032}}].
%%CITATION = HEP-TH 0103032;%%.

\bibitem{Hohm:2014qga}
O.~Hohm and H.~Samtleben, { Consistent {K}aluza-{K}lein truncations via
  exceptional field theory},  { JHEP} { 1501} (2015) 131,
[\href{http://xxx.lanl.gov/abs/1410.8145}{{\tt 1410.8145}}].
%%CITATION = ARXIV:1410.8145;%%.

\bibitem{deWit:2008ta}
B.~de~Wit, H.~Nicolai, and H.~Samtleben, { Gauged supergravities, tensor
  hierarchies, and {M}-theory},  { JHEP} { 0802} (2008) 044,
[\href{http://xxx.lanl.gov/abs/arXiv:0801.1294}{{\tt arXiv:0801.1294}}].
%%CITATION = ARXIV:0801.1294;%%.

\bibitem{Julia:1981wc}
B.~Julia, { Infinite {L}ie algebras in physics},  in { Johns Hopkins Workshop
  on Current Problems in Particle Theory}, 1981.

\bibitem{Julia:1982gx}
B.~Julia, { Kac-{M}oody symmetry of gravitation and supergravity theories},  in
  { Lectures in Applied Mathematics AMS-SIAM, Vol. 21}, p.~335, 1985.

\bibitem{Nicolai:1988jb}
H.~Nicolai and N.~Warner, { The structure of ${N=16}$ supergravity in two
  dimensions},  { Commun.Math.Phys.} { 125} (1989)
369.
%%CITATION = CMPHA,125,369;%%.

\bibitem{Damour:2002cu}
T.~Damour, M.~Henneaux, and H.~Nicolai, { {${E}_{10}$} and a `small tension
  expansion' of {M} theory},  { Phys. Rev. Lett.} { 89} (2002) 221601,
[\href{http://xxx.lanl.gov/abs/hep-th/0207267}{{\tt hep-th/0207267}}].
%%CITATION = HEP-TH/0207267;%%.

\bibitem{Damour:2006xu}
T.~Damour, A.~Kleinschmidt, and H.~Nicolai, { {$K(E_{10})$}, supergravity and
  fermions},  { JHEP} { 0608} (2006) 046,
[\href{http://xxx.lanl.gov/abs/hep-th/0606105}{{\tt hep-th/0606105}}].
%%CITATION = HEP-TH/0606105;%%.

\bibitem{West:2003fc}
P.~C. West, { {$E_{11}$}, {$SL(32)$} and central charges},  { Phys.Lett.} {
  B575} (2003) 333--342,
[\href{http://xxx.lanl.gov/abs/hep-th/0307098}{{\tt hep-th/0307098}}].
%%CITATION = HEP-TH/0307098;%%.

\bibitem{Peeters:2006kp}
K.~Peeters, { {A field-theory motivated approach to symbolic computer
  algebra}},  { Comput. Phys. Commun.} { 176} (2007) 550--558,
[\href{http://xxx.lanl.gov/abs/cs/0608005}{{\tt cs/0608005}}].
%%CITATION = CS/0608005;%%.

\bibitem{Peeters:2007wn}
K.~Peeters, { Introducing {C}adabra: {A} symbolic computer algebra system for
  field theory problems},
\href{http://xxx.lanl.gov/abs/hep-th/0701238}{{\tt hep-th/0701238}}.
%%CITATION = HEP-TH/0701238;%%.

\bibitem{Koepsell:1999uj}
K.~Koepsell, H.~Nicolai, and H.~Samtleben, { On the {Y}angian
  ${Y}(\mathfrak{e}_8)$ quantum symmetry of maximal supergravity in two
  dimensions},  { JHEP} { 04} (1999) 023,
[\href{http://xxx.lanl.gov/abs/hep-th/9903111}{{\tt hep-th/9903111}}].
%%CITATION = HEP-TH 9903111;%%.

\end{thebibliography}

\providecommand{\href}[2]{#2}\begingroup\raggedright\endgroup

\end{document}